\documentclass[preprint,review,10pt]{llncs}

\usepackage{graphicx}
\usepackage{longtable}
\usepackage{txfonts}
\usepackage{array} 
\usepackage{multicol,rotating}
\begin{document}
\title{Trans-Neptunian objects and Centaurs at thermal wavelengths}
\titlerunning{TNOs and Centaurs at thermal wavelengths}
\author{Thomas M{\"u}ller\inst{1}\orcidID{0000-0002-0717-0462} \and
Emmanuel Lellouch\inst{2}\orcidID{0000-0001-7168-1577} \and
Sonia Fornasier\inst{2}\orcidID{0000-0001-7678-3310}}
%
%\authorrunning{T. M{\"u}ller et al.}
\institute{Max-Planck-Institut f{\"u}r extraterrestrische Physik, Giessenbachstrasse 1, 85748 Garching, Germany; \email{tmueller@mpe.mpg.de} \and
LESIA, Observatoire de Paris, Universit\'e PSL, CNRS, Univ. Paris Diderot, Sorbonne Paris Cit\'{e}, Sorbonne Universit\'e, 5 Place J. Janssen, 92195 Meudon Pricipal Cedex, France}
%\email{tmueller@mpe.mpg.de}\\
%
\maketitle             % typeset the header of the contribution
%
%%%%%%%%%%%%%%%%%%%%%%%%%%%%%%%%%%%%%%%%%%%%%%%%%%
\begin{abstract}
{The thermal emission of transneptunian objects (TNO) and Centaurs has been observed at mid- and far-infrared wavelengths - with
the biggest contributions coming from the Spitzer and Herschel space observatories-, and the brightest ones also at sub-millimeter and millimeter wavelengths.
These measurements allowed to determine the sizes and albedos for almost 180 objects, and densities for about 25 multiple systems.
The derived very low thermal inertias show evidence for a decrease at large heliocentric distances and for high-albedo objects,
which indicates porous and low-conductivity surfaces. The radio emissivity was found to be low ($\epsilon_r$=0.70$\pm$0.13) with
possible spectral variations in a few cases. The general increase of density with object size points to different
formation locations or times. The mean albedos increase from about 5-6\% (Centaurs, Scattered-Disk Objects) to 15\% for the
Detached objects, with distinct cumulative albedo distributions for hot and cold classicals. The color-albedo separation in
our sample is evidence for a compositional discontinuity in the young Solar System. The median albedo of the sample (excluding dwarf
planets and the Haumea family) is 0.08, the albedo of Haumea family members is close to 0.5, best explained by the presence
of water ice. The existing thermal measurements remain a treasure trove
at times where the far-infrared regime is observationally not accessible.}

\end{abstract}
%%%%%%%%%%%%%%%%%%%%%%%%%%%%%%%%%%%%%%%%%%%%%%%%%%

%%%%%%%%%%%%%%%%%%%%%%%%%%%%%%%%%%%%%%%%%%%%%%%%%%
\section{Introduction}
%%%%%%%%%%%%%%%%%%%%%%%%%%%%%%%%%%%%%%%%%%%%%%%%%%

Thermal IR observations are crucial for the physical and thermal characterization
of distant objects which are typically too small to be resolved even by the largest
ground or space telescopes. The thermal emission measurements
allow one to determine accurate radiometric size and albedo information, but they
also put constraints on the object's thermal properties and spin-axis orientations.
Albedos derived from thermal data are important to interpret broad-band
colors and NIR spectral data.
The submm/mm-range emission originates from a few millimeters below the surface and
can be used to determine the object's long-wavelength emissivity.
A recent new approach is to combine multi-data information
(occultations, thermal, high-resolution imaging, etc.) to perform
more sophisticated physical and thermal characterization of TNOs and
Centaurs, and to constrain properties which are otherwise not
accessible, for example the object's spin-axis orientation. At the same time,
the multi-disciplinary approach allows
to improve and extend model techniques, e.g., the verification of
radiometric sizes via occultation results. A comparison between radiometric
and occultation size values is given in Ortiz et al.\ (same book).

However, the thermal emission from TNOs and Centaurs\footnote{We use the term
"trans-Neptunian object (TNO)" or "(Edgeworth-) Kuiper Belt object (KBO)"
to describe objects with orbits in the range 30 to 50\,AU, including also
the Scattered-disk objects (SDO) with high eccentricities and inclinations.
The Centaurs are closer and partly active objects with perihelia between
Jupiter and Neptune.} is difficult to detect, mainly
due to their large distance from the Sun (and the observer) and their
low surface temperatures\footnote{Blackbody radiation from TNOs and Centaurs peaks
in the FIR between 50 and 100\,$\mu$m, a range which is only accessible
from outside the Earth's atmosphere.}. 

We give an overview of the existing thermal emission measurements for TNOs and Centaurs
(Section~\ref{sec:sec1}), the modeling concepts (Section~\ref{sec:sec2}), and we present
the derived radiometric properties (Sections~\ref{sec:sec3} and \ref{sec:sec4}).
The chapter is concluded by a short outlook (Section~\ref{sec:outlook}).

%%%%%%%%%%%%%%%%%%%%%%%%%%%%%%%%%%%%%%%%%%%%%%%%%%
\section{Thermal data for TNOs and Centaurs}
\label{sec:sec1}
%%%%%%%%%%%%%%%%%%%%%%%%%%%%%%%%%%%%%%%%%%%%%%%%%%

The thermal infrared (IR) covers approximately the wavelength range from about 5 to 300\,$\mu$m,
often sub-divided into the mid-IR (MIR) range up to about 30\,$\mu$m (ground-based N-/Q-band,
SOFIA\footnote{\url{https://www.sofia.usra.edu/}},
WISE/NEOWISE\footnote{\url{https://neowise.ipac.caltech.edu/}},
AKARI\footnote{\url{https://www.ir.isas.jaxa.jp/AKARI/}},
Spitzer\footnote{\url{http://www.spitzer.caltech.edu/}}-IRS/IRAC)
and the far-IR (FIR) at wavelengths up to several hundred micron
(Herschel\footnote{\url{http://sci.esa.int/herschel/}}-PACS,
Spitzer-MIPS), followed by the sub-millimeter (submm) range below 1\,mm wavelength (e.g., Herschel-SPIRE;
CSO\footnote{\url{http://www.cso.caltech.edu/}};
ALMA\footnote{\url{http://www.almaobservatory.org/}} bands 7-10),
and the millimeter (mm) range (e.g., IRAM\footnote{\url{http://www.iram-institute.org/}},
ALMA bands 1-6, NOEMA\footnote{\url{http://iram-institute.org/EN/noema-project.php}}).

The first thermal measurements for a Centaur go back to
ground-based infrared and IRAS measurements of Chiron (Lebofsky et al. 1984; Sykes \&
Walker 1991; Campins et al. 1994), followed by millimeter observations
(Jewitt \& Luu 1992; Altenhoff \& Stumpff 1995). Pholus (Howell et al. 1992; 
Davies et al. 1993) and Chariklo (Jewitt \& Kalas 1998) were next, with 
solid detections mainly in Q-band at $\sim$20\,$\mu$m. A first overview of
the physical characteristics of TNOs and Centaurs (Davies 2000) listed sizes
for 8 Centaurs (3 based on thermal measurements, other sizes are estimated
assuming an albedo of 4\%) and no TNO sizes (Pluto was
still considered as planet at that time).

The Pluto-Charon system had the first thermal detections by IRAS in 1983
(Aumann \& Walker 1987, Sykes et al.\ 1987, Tedesco et al.\ 1987)
and a decade later at submm/mm (Stern et al.\ 1993, Jewitt 1994).
The characterisation of the system's IR and millimeter variability and
the corresponding modeling followed: Sykes (1999) using 
re-analysed IRAS data, Lellouch et al.\ (2000a) interpreting dedicated ISO-ISOPHOT
measurements, and Lellouch et al.\ (2000b) by searching for variability in
Pluto's milli\-meter-wave emission.
ISO also contributed with low-SNR FIR detections of two TNOs:
15789 (1993 SC) and 15874 (1996 TL66) (Thomas et al.\ 2000).
In the years after, Varuna was measured in the submm (Jewitt et al.\ 2001) and mm
(Lellouch et al.\ 2002), the Centaurs Asbolus and Chiron (Fern{\'a}ndez et al.\ 2002) were detected at
MIR, Chiron and Chariklo also at FIR/submm/mm (Altenhoff et al.\ 2001; Groussin et al.\ 2004),
the TNOs 55565 (2002 AW197) (Margot et al. 2002), 47171 Lempo (1999 TC36) (Altenhoff et al. 2004)
and Eris (Bertoldi et al.\ 2006) at radio wavelengths. However, many attempts to measure the TNO thermal emission
produced only upper limits (e.g.\ Altenhoff et al.\ 2004; Brown et al.\ 2004; Ortiz et al.\ 2004).
This changed dramatically with the availability of the Spitzer Space Telescope:
Stansberry et al.\ (2004) reported Spitzer-MIPS FIR detections of
14 KBOs and 8 Centaurs, Grundy et al.\ (2005) covered 20 TNOs, followed by several
other Spitzer-related TNO or Centaur projects (Cruikshank et al.\ 2005, 2006;
Stansberry et al.\ 2006; Grundy et al.\ 2007,2008). Stansberry et al.\ (2008) summarized
the Spitzer-provided constraints on the physical properties of 47 KBOs and Centaurs.
In the years after, the Spitzer-based studies focused either on individual TNOs
or smaller samples (e.g.\ Brucker et al.\ 2009; Stansberry et al.\ 2012).
Four Centaurs and seven TNOs were also detected with AKARI (T.\ Sekiguchi, priv.\ comm.),
however, no fluxes or radiometric results were published so far.

The next big step in thermal observations of Centaurs and TNOs came with Herschel's
large Open Time Key Project on "TNOs are Cool: A survey of the trans-Neptunian region
with Herschel" (M\"uller et al.\ 2009) which produced more than 20 publications. The 
Herschel data (partly also Spitzer data) allowed to interpret the thermal emission
of almost 170 TNOs and Centaurs:
M\"uller et al.\ 2010; Lellouch et al.\ 2010; Lim et al.\ 2010;
Barucci et al.\ 2012; Mommert et al.\ 2012; P\'al et al.\ 2012; Santos-Sanz et al.\ 2012; Vilenius et al.\ 2012;
Fornasier et al.\ 2013; Kiss et al.\ 2013; Lellouch et al.\ 2013;
Duffard et al.\ 2014; Lacerda et al.\ 2014; Vilenius et al.\ 2014;
Marton et al.\ 2015; P\'al et al.\ 2015;
Kiss et al.\ 2016; Lellouch et al.\ 2016; P\'al et al.\ 2016;
Kovalenko et al.\ 2017; Santos-Sanz et al.\ 2017;
Kiss et al.\ 2018; M\"uller et al.\ 2018; Vilenius et al.\ 2018.

%[HERE FIG.~\ref{fig:dr30}]
%\begin{figure}
%\includegraphics[width=0.9\textwidth]{DR30.png}
%\caption{WISE and Herschel-PACS measurements (color-corrected flux densities)
%         of the extreme solar system object 2012 DR30 observed at $\sim$14.5\,AU
%         distance, together with the best TPM solution. From Kiss et al.\ (2013).}
% \label{fig:dr30}
%\end{figure}

In parallel, the WISE project
performed an all-sky survey at MIR wavelengths and
detected the brightest Centaurs (Bauer et al.\ 2013).

Since Spitzer and Herschel, the situation has become much more difficult and only a dozen
TNOs have been observed in the submm/mm range, mainly by ALMA
(Moullet et al.\ 2011; Gerdes et al.\ 2017; Brown \& Butler 2017, 2018; Lellouch et al.\ 2017).

%%%%%%%%%%%%%%%%%%%%%%%%%%%%%%%%%%%%%%%%%%%%%%%%%%
\section{Radiometric techniques}
\label{sec:sec2}
%%%%%%%%%%%%%%%%%%%%%%%%%%%%%%%%%%%%%%%%%%%%%%%%%%

\subsection{Models to interpret thermal measurements}

Radiometry is a powerful technique to derive size, albedo, and thermal properties from thermal infrared measurements.
The technique consists in the exploitation and interpretation of thermal data (see Section~\ref{sec:sec1})
in combination with data in the visible. The visible data are mainly obtained from the ground and include the
object's absolute magnitude and estimates for the phase integral q.
%phase curve, usually described by the H-G parameterization (Bowell et al. 1989).
The most accurate radiometric properties are determined when objects are observed
close to their thermal emission peaks, preferentially shortwards and longwards of the peak.
For Centaurs and TNOs these peaks are located in the FIR regime between $\approx$50 and 100\,$\mu$m
(accessible from space only).
% But also shorter and longer wavelength observations are available (ground-based MIR or submm/mm
% measurements) and can be analysed by radiometric techniques.

There are different thermal models to obtain size and albedo of atmosphereless small bodies:
simple models like the Standard Thermal Model (STM), the Fast Rotating Thermal Model (FRM), or the near-Earth
asteroid thermal model (NEATM) or more complex thermophysical models (TPM) are used (see Delbo' et al. 2015 for a detailed
discussion). For Centaurs and TNOs, mostly the NEATM (Harris 1998) and the TPM (Lagerros 1996, 1997, 1998;
M\"uller \& Lagerros 1998, 2002) are applied. In cases where the object is lacking shape or spin properties, the NEATM
is often used. The beaming factor $\eta$ is then either fitted to multi-band thermal measurements or 
a default value is applied. The STM/NEATM concepts can also handle elongated shapes
(Brown 1985; Lellouch et al.\ 2017; Vilenius et al.\ 2018), but have severe limitations when
thermal measurements cover different aspect angles\footnote{The aspect angle is defined
such that it equals 0$^{\circ}$ when we observe the North pole and it equals 180$^{\circ}$
when we observe the South pole.}.

In cases where the rotation period is known,
% or when satellites or rings are present
the TPM concepts have the advantage that in addition to the standard size-albedo solution, also the spin-axis
and thermophysical parameters can be constrained (Fornasier et al.\ 2013; Kiss et al.\ 2018; M\"uller et al.\ 2018).
P\'al et al.\ (2012) applied STM and TPM techniques to the Herschel measurements of (90377) Sedna and 2010 EK139.
The derived size-albedo values agree within $\sim$10\%. P\'al et al.\ (2016) also show that NEATM (fitted
beaming factor) and TPM (assuming equator-on viewing geometry) approaches
lead to the same size-albedo conclusions for (225088) 2007 OR10.
% However, the 2007 OR10 case also showed that
However, the discovery of a satellite (Kiss et al.\ 2017) suggests a nearly pole-on viewing geometry for 2007 OR10.
The recalculated TPM size (Kiss et al.\ 2018) is 18\% smaller than the NEATM best-$\eta$ solution, but agrees
within 5\% when assuming a default $\eta =1.2$ in the radiometric calculations, indicating that
NEATM solutions require a proper error handling for the fits of the beaming factors.
%free parameter (P\'al et al.\ 2016).
% However, Kiss et al. (2018) was necessary after the discovery of a satellite (Kiss et al. 2017).

The TPM concept has advantages when objects have known spin or shape properties, or in cases
where multi-epoch thermal data for different observing geometries (like for Centaurs observed over
a wide range of heliocentric distances or aspect angles) are available (e.g.\ Kiss et al.\ 2013).
Recently, stellar occultations by Centaurs and TNOs revealed ring systems, elongated shapes, constraints
on possible atmospheres, and very accurate size information (see Ortiz et al., same volume). Here, the TPM 
applications allow to determine the object's thermal properties and testing of the spin-axis orientation
(Kiss et al.\ 2013; M\"uller et al.\ 2018). For multi-wavelength thermal data which include also submm/mm
observations, a wavelength-dependent emissivity within the TPM code is necessary to explain the observed fluxes
(Fornasier et al.\ 2013; Lellouch et al.\ 2017).

The accuracy of the radiometrically derived size and albedo values depends very much on
the model input values: a good-quality H-magnitude is needed for deriving the object's albedo.
A typical 0.1\,mag error for H translates into a $\sim$10\% error in geometric albedo.
The thermal flux error is crucial for the final size accuracy, but here the beaming
parameter (or thermal inertia and surface roughness in case of the TPM) and the object's
emissivity are also relevant. In general, the final quality of the radiometric size 
benefits from high-quality multi-epoch and multi-wavelengths observations. It is also important
to mention that radiometric solutions for high-albedo dwarf planets like Pluto, Eris, Makemake
suffer from unknowns for the object's surface scattering properties, expressed in the
phase integral (Stansberry et al. 2008). Using the formula $q = 0.336 p_V + 0.479$, with q being the
phase integral and p$_V$ the object's geometric V-band albedo, by Brucker et al. (2009)
solves only part of the problem since individual objects can deviate significantly
from that relation (Verbiscer et al. 2018). To account for a wide range of scattering properties for
icy surfaces, M\"uller et al. (2018) used a phase integral of $q = 0.65 \pm 0.20$ for Haumea
which has a geometric albedo of p$_V$ = 0.51 (Ortiz et al. 2017). For higher-albedo objects the uncertainty in q might
be even larger, for lower-albedo objects the phase integral is less relevant and a
default value of 0.39 (Bowell et al. 1989) can be used.

\subsection{Satellite thermal emission}

Almost all large TNOs have satellites (Parker et al.\ 2016b; Kiss et al.\ 2017) and these satellites
can contribute to the observed thermal emission measurements. At MIR and FIR it was so far not
possible to resolve the satellite and the main-body emission with the current technical limitations.
Only in the submm/mm range it is possible to detect large satellites around dwarf planets with 
the high spatial resolution of the ALMA array. Brown \& Butler (2018) examined the spatially resolved
Orcus-Vanth and Eris-Dysnomia systems and found low albedos similar to other
TNOs of similar size and different from the main body's properties. It is therefore reasonable
to estimate the thermal emission of satellites by using adequate beaming and albedo properties
from published statistical analysis of TNO samples.
However, the thermal emission estimate for satellites remains uncertain. This has to be
considered in the radiometric analysis in cases where primaries and secondaries
cannot be seen directly (e.g., for 2007 OR10: see discussion in Kiss et al. 2018;
for Haumea: see discussion in M\"uller et al. 2018). 

\subsection{Ring thermal emission}

Rings have been detected around Chariklo (Braga-Ribas et al.\ 2014) and Haumea (Ortiz et al.\ 2017) and are suspected
on other bodies (e.g. Ortiz et al.\ 2015).  Lellouch et al.\ (2017) estimated the contribution of
rings to thermal emission, using a simplified version of a model for Saturn's rings. In this model, the only source of energy for
ring particles is absorbed solar radiation, but mutual shadowing -- as seen both from the Sun and the observer -- and
optical depth effects are taken into consideration. The model further assumes that ring particles have a bolometric and
spectral emissivity of unity. By analogy with Saturn's rings, the latter assumption is likely valid up to $\sim$200 $\mu$m,
but the spectral emissivity could decrease at longer wavelengths if ring particles are made of water ice. Model free
parameters are the ring radius, width, opacity, and Bond albedo (related to the I/F reflectivity). The relative importance
of ring contribution to solid body emission is of course strongly geometry-dependent, but was found to have a minor
effect, e.g. affecting the equivalent diameter determination by at most 5 \% for Chariklo. In that sense,  for ellipsoid bodies,
including/omitting rings is much less important than properly accounting for the varying pole orientation and hence
projected surface of the object.

%\section{TNOs physical properties and results}
%%%%%%%%%%%%%%%%%%%%%%%%%%%%%%%%%%%%%%%%%%%%%%%%%%
\section{Albedos, sizes and densities}
\label{sec:sec3}
\begin{figure}
\includegraphics[width=0.99\textwidth]{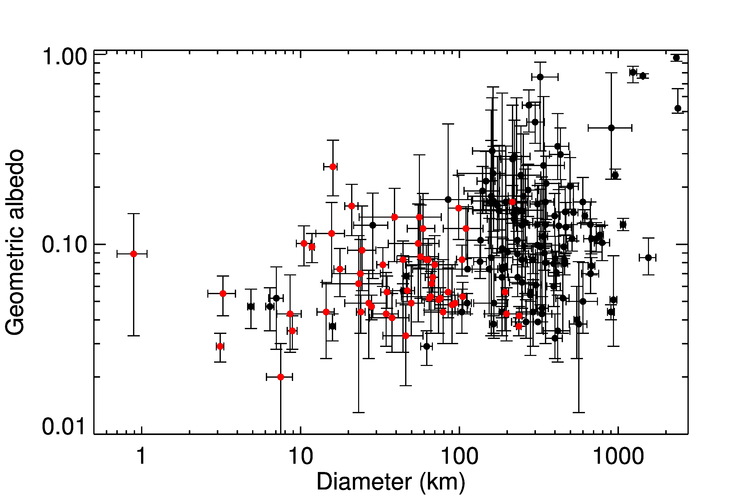}
\includegraphics[width=0.99\textwidth]{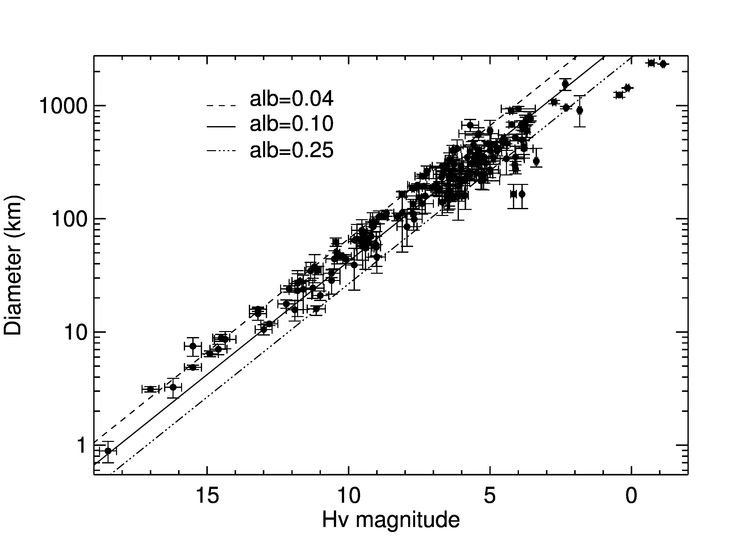}
\caption{Top: Geometric albedo versus diameter for 170 TNOs (in black) and Centaurs (in red) from Herschel, Spitzer, and WISE data
(see Table~\ref{tbl:pv_all}, but excluding objects with upper/lower limit estimates).
The selection bias due to distance is clearly visible as small sizes are measured only for closer bodies, i.e.\ Centaurs.
Bottom: Diameter versus H$_v$ magnitude for 170 TNOs and Centaurs. Absolute magnitude with an albedo of 10\% is a good
proxy for size for the majority of TNOs.}
 \label{fig1}
\end{figure}

%%%%%%%%%%%%%%%%%%%%%%%%%%%%%%%%%%%%%%%%%%%%%%%%%%
[HERE FIGURE ~\ref{fig1}]

Sizes, albedos, and in many cases beaming factors, derived from Spitzer, WISE and Herschel thermal data,
are available for 178 TNOs and Centaurs
and listed in Table~\ref{tbl:pv_all}. Transneptunian bodies show a huge diversity both in albedo and size (Fig.~\ref{fig1}, left side). 
Measured diameters range from a few tens of km to $\sim$2000\,km
but the distribution of sizes is strongly affected by discovery and selection biases in the Spitzer/Herschel/WISE sample. 
Eleven objects have diameters larger than 900\,km: Charon, Orcus, Quaoar, Salacia, 2002 MS$_4$, 2007 OR$_{10}$, Sedna and the 4 so-called dwarf planets Pluto, Eris, Makemake and Haumea. Geometric albedos $p_V$ vary over a factor of $\sim$25-30 from  $p_V = 3-4$\% for
the darkest objects to $p_V \sim$ 50-90\% for the brightest (see Table~\ref{tbl:pv_all}). 
The latter category includes 1) volatile-rich bodies (Eris, Makemake, Pluto), where the high albedo likely results from seasonally-variable
resurfacing processes (deposition of fresh ice); 2) Haumea and some of its collisional family members, covered by pure H$_2$O ice likely
excavated from Haumea's mantle.
Excluding these bodies,
the vast majority of TNOs have albedos ranging from 4 to 25 \%, with a median (mean \& standard deviation) of 8\% (10 $\pm$ 6\%), see Table~\ref{tbl:pv}.
As routinely assumed, H$_v$ magnitude is a good proxy for size (Fig.~\ref{fig1}, right side): using a 10 \%
albedo ensures an error on the diameter lower than a factor of 1.6. Fig.~\ref{fig1} (left side) further shows that except for the very bright objects,
there is no obvious correlation between albedo and diameter, the only trend being (except for one outlier, 2005 UJ$_{438}$, whose radiometric
solution may be flawed by coma activity) the lack of objects with albedos $>$ 0.2 at diameters $<$ 100\,km.
In Table~\ref{tbl:pv} we present the median, mean, standard deviation $\sigma$, the error of the mean ($\sigma/\sqrt{n}$), and
the minimum and maximum albedo values for each dynamic group. In the calculations we excluded the objects which have only
upper/lower-limit radiometric solutions. For the Hot Classicals and the Detached categories, we present these values with and without
the dwarf planets and the Haumea family. We also looked at the weighted mean albedos which are typically below the median
values. But the albedo errors are closely connected to the H-magnitude errors which are coming from different sources, including
rough estimates or affected by unknown opposition effects and different handling of lightcurve amplitudes. It is therefore problematic
to work with error-weighted albedos.

\begin{figure}
\centering
\includegraphics[width=0.8\textwidth]{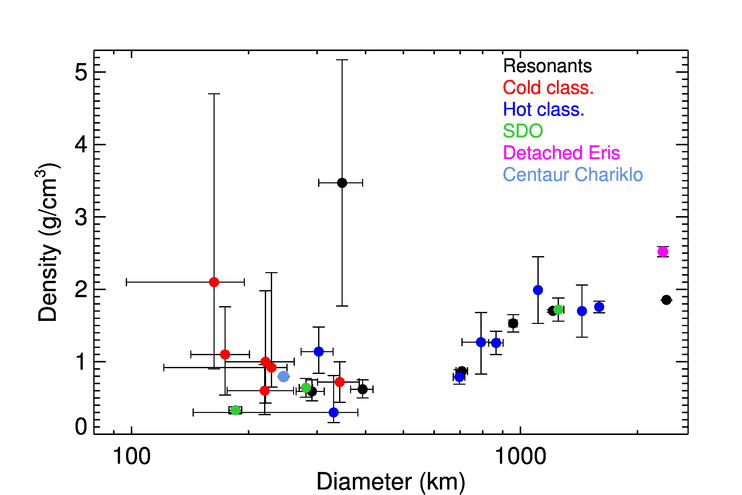}
\caption{Bulk density versus the effective diameter for the observed TNOs. Red symbols - cold classicals; blue symbols - hot Classicals; black symbols - resonants; green symbols - SDO; magenta symbol-detached. Density values are from the references given in the text, as well as a description of the outliers. We add also Pluto \& Charon from New Horizons mission results.}
 \label{density}
\end{figure}

Twenty-six binary systems with known system masses have equivalent diameter values, yielding density estimates.
Density values are reported in
Sicardy et al.\ (2011), Mommert et al.\ (2012), Ortiz et al.\ (2012, 2017), Santos-Sanz et al.\ (2012), Vilenius et al.\ (2012, 2014),
Brown (2013), Fornasier et al.\ (2013), Lellouch et al.\ (2013),
Brown \& Butler (2017), Dias-Oliveira et al.\ (2017), Kovalenko et al.\ (2017), Leiva et al.\ (2017),
Kiss et al.\ (2018), and Stern et al.\ (2018).
The derived bulk densities span a wide range, from below that of water ice to that of nearly pure rock (Fig.~\ref{density}).
TNOs smaller than 400\,km have density lower than 1 g/cm$^3$, as already noticed (for 1999 TC$_{36}$; Stansberry et al.\ 2006), and this implies both a small rock-to-ice ratio and a high porosity. Possible outliers are the cold classical Borasisi and the 2:1 resonant body 2002 WC$_{19}$; those may have higher densities but uncertainties are huge. For comparison, precise measurement of the 67P/Churyumov Gerasimenko nucleus by the Rosetta mission yields a density of 537.8$\pm$0.7 kg/m$^{3}$, and a large porosity (70-80\%, Sierks et al.\ 2015; P\"atzold et al.\ 2016; Preusker et al.\ 2017).
Vilenius et al. (2014) also found that the sizes of the binaries components for objects smaller than 400\,km are not significantly different from each other.\\ 

[HERE FIGURE ~\ref{density}]

The general increase of density with object size (Fig.~\ref{density}) may be partly explained by gravitational self-compaction, reducing the macroporosity. However, this
process alone seems unlikely to explain the factor-of-five variation of density from small to large bodies, so that large KBOs may truly be rock-richer than small ones.
Thus accreting large rock-rich KBOs from smaller, rock-poor bodies is not a viable scenario (Brown 2013).
Dwarf planets may have larger rock-to-ice ratios possibly due to different formation location/times compared to smaller ones.
Barr \& Schwamb (2016) note that among dwarf planets, those with large moons tend to
have smaller densities than those with small moons\footnote{However, Fig.~2 in Barr \& Schwamb (2016)
suggests that the correlation between density and satellite mass ratio is not strong.}.
They propose that low-velocity collisions between undifferentiated primordial dwarf
planets make large planet/moon pairs, in which both bodies retain
their original compositions, while higher velocity collisions between differentiated dwarf planets could yield
rock-enriched primaries with small ice-rich satellites.

The median albedos increase from $\sim$5-6\% for Centaurs and Scattering Disk objects (SDO) to $\sim$8-10\% for Hot Classicals and Plutinos,
to $\sim$14\% for Cold Classicals and to $\gtrsim$15\% for Detached objects (see Table~\ref{tbl:pv}).
Within the Classicals, the cold and hot populations have a clearly distinct
albedo cumulative probability distributions.
Given the clustering of objects in two groups in the color-albedo diagram 
(Lacerda et al.\ 2014), and the hypothesis that colors reflect the 
formation distance (Brown et al.\ 2011), this indicates that cold and 
hot classicals likely formed in different regions of the Solar System.
Particularly remarkable is the similarity of the albedo distribution of Centaurs and SDO (median p$_V$ = 5-6\%, Lacerda et al.\ 2014,
and Table~\ref{tbl:pv}).
It is consistent with an origin of Centaurs in the Scattered Disk
(Volk \& Malhotra 2008), and may suggest that these transitioning objects do not change surface properties when entering the Giant Planet region,
although Jewitt (2009) and Melita \& Licandro (2012) have advocated that the color bimodality of Centaurs is caused by differential thermal processing.

\begin{table}
\caption{Average albedo values for different dynamic groups: the number of objects
in a given sample (n), the median albedo value, the mean and standard deviation ($\sigma$),
the standard error of the mean ($\sigma/\sqrt{n}$),
the minimum and maximum values (min/max) in the sample, and comments on the sample: "all" refers
to all entries in Table~\ref{tbl:pv_all} with diameter and albedo solutions, excluding the
objects with only upper/lower limit solutions. In cases of dwarf planets within a given dynamic
group we list the values with and without these special objects.
\label{tbl:pv}}
\begin{tabular}{rrcccccl}
\noalign{\smallskip}
\hline
\noalign{\smallskip}
Dynamic group    & n  & median & mean & $\sigma$ & $\sigma/\sqrt{n}$ & min/max & comment \\
\noalign{\smallskip}
\hline
\noalign{\smallskip}
Cold Classical   & 17 & 0.136 & 0.132 & 0.045 & 0.0109 & 0.054 0.236 & all \\                                   %TM_final: weighted mean: 0.099
Hot Classical    & 32 & 0.107 & 0.195 & 0.225 & 0.0397 & 0.032 0.804 & all \\                                   %TM_final: weighted mean: 0.073
                 & 26 & 0.084 & 0.102 & 0.072 & 0.0141 & 0.032 0.310 & without Makemake, Haumea \& family \\    %TM_final: weighted mean: 0.060
Plutino          & 25 & 0.089 & 0.114 & 0.064 & 0.0128 & 0.039 0.281 & without Pluto/Charon \\                  %TM_final: weighted mean: 0.093
Outer Resonant   & 11 & 0.163 & 0.148 & 0.071 & 0.0215 & 0.049 0.297 & all$^{1}$ \\                             %TM_final: weighted_mean: 0.100
Detached         &  8 & 0.167 & 0.291 & 0.294 & 0.1040 & 0.079 0.960 & all \\                                   %TM_final: weighted_mean: 0.132
                 &  7 & 0.148 & 0.195 & 0.125 & 0.0471 & 0.079 0.410 & without Eris \\                          %TM_final: weighted mean: 0.091
SDO              & 20 & 0.057 & 0.075 & 0.046 & 0.0103 & 0.037 0.231 & all \\                                   %TM_final: weighted_mean: 0.046
Haumea family    &  9 & \multicolumn{5}{c}{median 0.48$^{+0.28}_{-0.18}$} &  Vilenius et al.\ 2018 \\           %
Centaur          & 55 & 0.056 & 0.074 & 0.043 & 0.0058 & 0.020 0.256 & all \\                                   %TM_final: weighted mean: 0.046
\noalign{\smallskip}
\hline
\noalign{\smallskip}
All TNOs \& Centaurs & 170 & 0.083 & 0.126 & 0.138 & 0.0106 & 0.020 0.960 & all \\                              %TM_final: weighted mean: 0.057
                     & 160 & 0.083 & 0.099 & 0.062 & 0.0049 & 0.020 0.328 & without P/C,E,M,S,H,Hfam$^{2}$ \\   %TM_final, excluding P/C,E,M,S,H,Hfam: weighted mean: 0.055
\noalign{\smallskip}
\hline
\noalign{\smallskip}
\end{tabular}
$^{1}$ : 2001 YH140 is in a 3:5 resonance with Neptune and not considered here;\\
$^{2}$ : Pluto/Charon, Eris, Makemake, Sedna, Haumea \& Haumea family members.
\end{table}

\subsection{Classical population}

Classicals - as well as the other dynamical classes - are defined according to the Gladman et al.\ (2008) dynamical classification.
Vilenius et al. (2012, 2014) analyzed 44 classical TNOs, whose sizes range from $\sim$ 130 to 930\,km, from the Herschel and Spitzer thermal data. 
Classicals are distinguished between the cold and hot populations according to inclination with a limit at  5$^{\circ}$. \\
Vilenius and co-workers found that the cold and hot classical populations are distinct in terms of size and albedo.
First, they have different averaged albedos, as previously noticed (Grundy et al.\ 2005; Brucker et al.\ 2009):
the cold classicals have a median albedo of 14\%
(mean: 13 $\pm$ 5\% on a sample of 17 objects),
higher than that of the hot population with 8\% (mean 10 $\pm$ 7\% from a sample of 26 objects),
excluding the Haumea family members. Second, they have different (debiased) size distributions: cold classicals are smaller than 400\,km, while the hot ones have a wider size distribution, and diameters up to $\sim$ 900\,km (excluding Makemake). The cumulative size distribution is also clearly different, having a much steeper slope (q=5.1$\pm$1.1) for cold  than for hot (q=2.3$\pm$0.1) classicals, evaluated for $160<D<280$\,km and $100<D<500$\,km, respectively.
Similarly, Fraser et al.\ (2014) found that dynamically quiescent ("cold") KBOs have a
steeper bright-end slope than excited ("hot") KBOs, although the distinction vanishes in the faint
end of the distribution. \\ 
The fact that the cold and hot populations are distinct in terms of spectral slope (Tegler \&
Romanishin, 2000; Doressoundiram et al.\ 2008), number of binary systems (Noll et al.\ 2008
and this book), size and albedo (Vilenius et al.\ 2014), strengths the hypothesis that these
two sub-populations had different dynamical histories and therefore formed in different
regions of the planetary disk. These results are consistent with the
Nice model (Batygin et al.\ 2011), indicating that  hot Classicals formed closer to the Sun
compared to their present location and migrated during the early Solar System evolution,
while the cold population formed in-situ.

\subsection{Resonant}

Thirty-seven resonant TNOs were investigated in the thermal wavelengths. Most of them (25) are Plutinos populating the 3:2 resonance with Neptune (Table~\ref{tbl:pv_all}).
The Plutinos sizes range from  about 85\,km to 730\,km  for the largest body investigated (2003 AZ84), except the dwarf planet Pluto.
Their albedos range from 4 to 28\%, as reported in Mommert et al.\ (2012) and Lellouch et al.\ (2013), with a median (mean) value p$_V$= 9\% (11 $\pm$ 6\%)
comparable to that of Centaurs, Jupiter family comets and other TNOs, excepting the detached and cold classicals. Mommert et al. (2012) found that
the Plutino size distribution is reproduced using a cumulative power law with q = 2 at sizes ranging from 120-400\,km and q = 3 at larger sizes. They did not find any
correlation between the different physical parameters of the Plutinos, except a clear anti-correlation between eccentricity and diameter,
likely caused by a discovery bias. Among the 18 Plutinos investigated in their study, six have evidence of water ice and show an higher albedo ($>$ 11\%) than the average Plutino value.
\\
The investigated outer resonants are brighter ($p_V$=16\% median and 15$\pm$7\% mean values) and bigger than Plutinos, very likely because of the
discovery and selection biases. Their size ranges from $\sim$ 150\,km to $\sim$ 1200\,km for the largest body investigated,
2007 OR10 (P\'al et al.\ 2016; Kiss et al.\ 2018), located in the 3:10 mean motion resonance with Neptune.

\subsection{Detached/SDO population}

The thermal properties were determined for 20 Scattered disk objects (SDO) and 8 Detached transneptunians (Table~\ref{tbl:pv_all}). Santos-Sanz et al. (2012) analyzed 15 scattered-disk/detached objects. They report that SDO are smaller than 400\,km (excluding 2007 OR10, which is also classified as outer resonant), while detached objects are larger than 250\,km, the larger size of the measured detached bodies being again a discovery/selection bias. Excluding the dwarf planet Eris, the two populations have
different median (mean) albedos: 6\% (8 $\pm$ 5\%) for the SDOs,
and 15\% (20 $\pm$ 13\%) for the detached bodies. Santos-Sanz et al.\ (2012) and Stansberry et al.\ (2008) proposed that larger objects have
higher albedo because they may more easily retain volatiles. A possible positive correlation between albedo, size and perihelion distance was interpreted as a consequence of 
increased volatile sublimation and/or space weathering effects at low heliocentric distances, both of which leading to surface darkening.

\subsection{Centaurs}

Size, albedo and thermal properties were derived for 55 Centaurs combining the Spitzer,
Herschel (and in some cases WISE) observations
(Bauer et al.\ 2013; Duffard et al.\ 2014; Tegler et al.\ 2016; Romanishin \& Tegler 2018).
Centaurs have a median (mean) albedo of 6\% (7 $\pm$ 4\%).
Most of the Centaurs (92\%) are smaller than $\sim$120\,km in diameter, and they are thought to be fragments
from collisions of larger parent bodies (Pan \& Sari 2005).
Regarding sizes, Duffard et al.\ (2014) claim a lack of objects with sizes between 120 and 190\,km, 
while the largest two, 2002 GZ$_{32}$ and Chariklo, reach D $\sim$ 240\,km. They further report
a size dichotomy between red Centaurs, which are all small ($<$120\,km) and grey ones,
which can reach $\sim$240\,km.

\subsection{Haumea family}

Haumea is in a weak 7:12 resonance with Neptune (Ragozzinee \& Brown 2007) and it is hypothesized to be
the parent body of a TNO collisional family (Brown et al. 2007). Vilenius et al. (2018) found
that the thermally detected Haumea family members have high albedos in the range $\sim$0.3-0.8
(median p$_V$ = 0.48$^{+0.28}_{-0.18}$), indicative for the presence of water ice. 
Deep water ice signatures are actually detected spectroscpically on Haumea, one of its moons
(Hi'iaka) and several of the family members (see Brown et al.\ 2012).
Also the cumulative size distribution for sizes in the range 175-300\,km is
steeper (q=3.2$^{+0.7}_{-0.4}$) than for dynamical interlopers\footnote{
Interlopers (as defined by Ragozzine \& Brown 2007) belong to the
same dynamical cluster as family members but lack the H$_2$O spectral features.}
with D$<$500\,km.

\subsection{Colors/albedo correlations}

A comprehensive study of the color-albedo distribution for 109 TNOs and Centaurs with available visible spectral slopes or colors
and albedo values (see Table~\ref{tbl:pv_all}) was presented by Lacerda et al.\ (2014). They found that excluding the dwarf planet and the Haumea family members, the Transneptunian population is globallly split in two clusters: the dark neutral and the bright red objects. The dark neutral group  include spectrally gray objects, having an average spectral slope $S \sim$ 10\%/100 nm and values $<$ 18 \%/100 nm, and low albedo values, clustered at p$_V \sim$ 5\%. The red and bright objects have $18 < S < 58 $ \%/100nm, and albedos $>$ 6\%, with a higher median albedo value (around 15\%).  The dwarf planets and Haumea family members represent a third group of objects, having high albedo and small spectral slope values resulting from the presence of surface volatiles or exposed water ice. \\

The detailed analysis of spectral slope versus albedo for the different dynamical populations shows that Centaurs, Plutinos, hot Classicals and SDOs have objects in both groups, while cold Classicals, detached and outer resonants have only bright red TNOs (Lacerda et al.\ 2014). These results agree with conclusions from 
previous studies (that lacked the albedo information), finding that dynamically excited TNO populations are composed of
two main types of surfaces (Fraser \& Brown 2012, Bauer  et  al.\ 2013). This agrees with the scenario
(Malhotra 1995; Gladman et al.\ 2002; Batygin et al.\ 2011) that cold Classicals, detached and outer resonants
formed in distant regions from the Sun and that their bright red surfaces may be related to
heliocentric-distance-dependent fractionation of surface volatiles with different sublimation temperatures
(Brown et al.\ 2011; Fraser \& Brown 2012; Wong \& Brown 2016), with methanol
and/or hydrogen sulfide playing a key role\footnote{We note that CH$_3$OH has been detected
on a few objects, but H$_2$S has not so far (see chapter by Barucci and Merlin).}.
Conversely, hot Classicals, Plutinos, Centaurs and SDOs, formed closer to the Sun and in a wider range of heliocentric distances,
from 20 AU to about 48 AU (Petit et al.\ 2011), resulting in a wider range of physical properties,
both in term of surface colors and albedo.

The color-albedo splitting in two groups is particularly prominent for Centaurs (Bauer et al.\ 2013, Duffard et al.\ 2014,
Tegler et al.\ 2016, Romanishin \& Tegler 2018) with median albedos of $\sim$5 \% for the dark-neutral group and $\sim$8.4\%
for the bright red, with the red group also having lower mean inclinations.
An early explanation of the Centaur color bimodality in terms of collisional resurfacing has been invalidated by Th\'ebault \&
Doressoundiram (2003). Thermal processing has also been invoked (Melita \& Licandro 2012) whereby red objects tend to
spend shorter amount of times at small heliocentric distances; but in this case, Pholus would be a striking counter example.
Romanishin \& Tegler (2018)  find
that grey Centaurs have albedos not significantly different from those of the Trojans, consistent with a common origin.

%%%%%%%%%%%%%%%%%%%%%%%%%%%%%%%%%%%%%%%%%%%%%%%%%%
\section{Thermal and emissivity properties}
\label{sec:sec4}
%%%%%%%%%%%%%%%%%%%%%%%%%%%%%%%%%%%%%%%%%%%%%%%%%%

\subsection{Thermal inertia}
In addition to size, shape and albedo, physical properties accessible to thermal measurements include thermal inertia, surface roughness and surface emissivity.
Thermal inertia $\Gamma$ is related to thermal conductivity $\kappa$ through a relation that also involves the material density $\rho$ and
heat capacity $C$ ($\Gamma$ = $\sqrt {\kappa\rho C}$). The latter two parameters are usually unknown but as pointed out by Delb\'o et al. (2015), their
plausible range of variation is much smaller than that for $\kappa$ -- which can vary by order of magnitudes between fine grain regolith and
compact rock/ices. This justifies that $\Gamma$, the most directly accessible variable in a thermophysical model, be used as a proxy for $\kappa$,
itself an indicator on the processes contributing to thermal conduction (intragrain, intergrain, or radiation-assisted heat transfer,
e.g.\ Gundlach \& Blum 2013; Ferrari \& Lucas 2016). 

\subsubsection{Ensemble properties}

The unambiguous determination of an object's thermal inertia requires good-quality spin- and shape information
(e.g.\ Hanu\v{s} et al.\ 2018). Rotation periods have been measured for $\sim$150 objects, but less than 10\% of them have shapes
and pole direction available. Therefore, the simple NEATM model still remains the default approach, when dealing with ``large" samples.
Although the typical NEATM accuracy is 15 \% on diameter (Wolters \& Green 2009), NEATM-based biases for asteroids occur mostly at
large phase angle, a problem that is not relevant for TNOs/Centaurs. The NEATM-derived diameters usually agree with equivalent
diameters from stellar occultation values to within much better accuracy (see Ortiz et al., same book). The beaming factor is then
a proxy for the combined effect of thermal inertias, spin state and surface roughness. In rare cases where the thermal
inertia is estimated, it characterizes the response of the object to the diurnal cycle and pertains to a
layer comparable to the diurnal skin depth.

Lellouch et al.\ (2013) compiled (or re-rederived) 85 beaming factors ($\eta$) values for TNOs/Centaurs to assess ensemble thermal inertia properties
for the population. They found that: (i) beaming factors range from values $<$ 1 to $\sim$2.5 (close to the maximum value expected in the limit
of high thermal inertia for an equator-on object); (ii) $\eta$ values $>$2 are lacking at small ($<$ 30 AU) heliocentric distances;
(iii) beaming factors lower than 1 occur frequently ($\sim$ 40 \% of the objects), implying that surface roughness effects are important. 
Based on a statistical description of the bodies' spin properties and surface roughness, they determined that all these trends
could be explained by a mean thermal inertia of $\Gamma$ = (2.5$\pm$0.5) J m$^{-2}$s$^{-0.5}$K$^{-1}$ (SI),
with evidence for a decrease of $\Gamma$ with increasing heliocentric distance $r_h$, corresponding to a power law
$\Gamma$ $\sim$ $r_h^{-(1.0~to~1.7)}$. Another finding was that high-albedo objects have preferentially lower thermal inertias. 

Such thermal inertias are considerably smaller than for other Solar System icy bodies (e.g. $\Gamma$ = 50-70 for Jupiter's satellites, 5-20 SI for 
Saturn's (Howett et al.\ 2010 and references therein),  $\Gamma$ = 10-25 for Pluto/Charon, see below). They are also 2-3 orders of magnitude smaller than values
for compact ices, typically 2000-2600 SI (resp. 700-160 SI) for crystalline (resp. amorphous) ice at 150-30 K. Surface porosity and granularity strongly reduce the thermal
conductivity, and are likely to be the cause of the measured low thermal inertias. Regarding the temperature dependence of the thermal inertia,
Lellouch et al.\ (2013) noted that if thermal conduction is dominantly radiation-assisted (i.e. $\kappa \propto $ T$^3$), $\Gamma$ is 
expected to be nearly proportional to T$^2$, i.e. $r_h^{-1}$. Based on a model of grain contact and radiative conduction, Ferrari \& Lucas (2016)  
could quantitatively explain the low $\Gamma$ values and their heliocentric distance if amorphous ice is present at cm depths below a thin layer of crystalline ice.
The smaller value of $\Gamma$ for "the mean TNO" compared e.g. to Pluto/Charon does not necessarily imply different regolith properties. Rather it may be related
to different diurnal skin depths, which for equal density / thermal properties, is typically 4-5 times shallower on a 8 h-rotating TNO
than on the 6.39 day period Pluto/Charon. This interpretation implies that the thermal inertia increases with depth, which is confirmed
by recent estimates of the seasonal thermal inertia (Bertrand \& Forget 2016).

\subsubsection{Pluto/Charon and other prominent TNOs}
Thermal inertias of Pluto ($\Gamma$ = 16-26 SI) and Charon ($\Gamma$ = 9-14 SI) have been estimated based on unresolved
thermal lightcurves of the system, observed by ISO, Spitzer and Herschel (Lellouch et al.\ 2000a, 2011, 2016), covering altogether 20-500 $\mu$m.
The pre-New Horizons Pluto model was described with three surface units (N$_2$ ice, CH$_4$ ice, and H$_2$O-tholin) and the determined
thermal inertia globally applies to the non-N$_2$ ice (i.e. the non-isothermal) regions. The spatially-separated measurements of the cm/mm/submm
brightness temperature (T$_B$ = 31-33 K for Pluto vs 42-45 K for Charon; Butler et al.\ 2015 and personnel communication) are consistent with estimated
surface temperatures (Lellouch et al.\ 2011). The seemingly
higher thermal inertia for Pluto vs Charon may result from a significant contribution of atmospheric conduction
within a porous upper surface (Spencer \& Moore 1992, Lellouch et al.\ 2000b).

At Pluto, the seasonal skin depth is 120 times larger than the diurnal skin depth for a given
set of thermal properties. Determining the seasonal thermal inertia requires
measurements over seasonal timescales, or at least a temperature value in the polar night
(Leyrat et al.\ 2016).
But an object's sub-surface seasonal inertia also controls volatile cycles and the distribution of ices
at a point in time (e.g.\ Young 2013).
Bertrand \& Forget (2016) found that Pluto's pressure cycle, as well as the accumulation of
N$_2$ and CO ices in the low-latitude Sputnik Planitia basin, can be explained by a seasonal
thermal inertia $\Gamma_{seasonal}$ = (500-1500) SI, with 800 SI as the prefered value,
corresponding to an annual skin depth of $\sim$40\,m.

Although Charon lacks an atmosphere, a somewhat similar reasoning may be used to constrain its seasonal thermal inertia. Grundy et al. (2016) interpret
the red coloration of Charon's poles as due to winter cold-trapping of methane gas escaping from Pluto's atmosphere and captured by Charon, followed
by photolytic processing of the material into more complex and less volatile molecules. For methane to be cold-trapped, Charon's winter poles temperatures
must fall below an estimated 25 K. This condition is met for a seasonal thermal inertia $\Gamma_{seasonal}$ =  2.5-40 SI (i.e. comparable to the diurnal value),
but conversely it implies some upper limit to $\Gamma_{seasonal}$, and it is likely that a value as high as the Pluto one would violate the constraint.
These results  (i) confirm the vertical variation of Pluto's thermal inertia with depth (ii) suggest that Charon's seasonal thermal inertia is
smaller than Pluto's, as is the case for the diurnal $\Gamma$.

For more details on the Pluto-Charon system and the interpretation of New Horizons measurements see also the 
chapters by Grundy et al.\ and Spencer et al.\ (same book).

Dwarf planet (136138) Haumea is the only other TNO with a definitely detected thermal lightcurve (Lellouch et al.\ 2010, Lockwood et al.\ 2014,
Santos-Sanz et al.\ 2017; M\"uller et al.\ 2018). Shape information is also available from high-quality lightcurve observations (e.g.\ Lacerda
et al.\ 2008), and most
recently from stellar occultation, which also provides the pole orientation from the detection of a ring system (Ortiz et al.\ 2017).  
Adopting a shape model from Lacerda
et al.\ (2008), Santos-Sanz et al.\ (2017) inferred an extremely small thermal inertia ($\Gamma$ $<$ 0.5 SI), essentially constrained by the lack of temporal shift
between the optical and thermal lightcurves, and modelling the lightcurve amplitude then required a phase integral $q$ $>$ 0.73. After the occultation results, M\"uller et al.\ (2018) revisited
the problem by (i) re-reducing the Herschel data; (ii) using the occultation-lightcurve derived 3-D size-spin-shape solution;
(iii) estimating and correcting for the contribution of satellites and rings to the thermal flux. They reported a thermal inertia in the range
2-15 SI, with $\Gamma$~=~5 SI as the most likely value. The higher value compared to the previous estimate results from the combination
of (i) a strongly subdued lightcurve amplitude in the latest version of data reduction; (ii) the larger effective diameter and lower albedo yielded by the 
occultation results; (iii) the neglect of satellite/rings fluxes in the former study. In spite of the large difference, the new value for 
$\Gamma$ is in light with the gross picture of low thermal inertia for the TNO population. Further reanalysis for Haumea might still be warranted in the light
of the newly available value for $q$ (0.45 from Verbiscer et al.\ 2018).
 
Thermal inertia results for other individual TNOs, derived from thermophysical models exploring parameter space (diameter, thermal inertia, roughness, spin state), 
are scarce. They are summarized in Table~\ref{tbl:ti}.

\begin{center}
\begin{table}
\caption{Thermal inertia values for individual TNOs.
\label{tbl:ti}}
\begin{tabular}{lc}
\noalign{\smallskip}
\hline 
\noalign{\smallskip}
Object           & Thermal inertia (SI) \\
\noalign{\smallskip}
\hline 
\noalign{\smallskip}
Chiron           & 3-10 $^{(1)}$, 0-10 $^{(2)}$ \\
Chariko          & 3-30 $^{(1)}$, 1.5-10 $^{(2)}$\\
Quaoar           & 2-10 $^{(1)}$ \\
Orcus            & 0.4-2 $^{(1)}$ \\
Bienor           & 6-20 $^{(2)}$\\
2007 UK$_{126}$  & 0.7-10 $^{(3)}$ \\
2013 AZ$_{60}$   & $\gtrsim$50 $^{(6)}$ \\
2007 OR$_{10}$   & 1-10 $^{(7)}$ \\
Haumea           & 0-0.5 $^{(4)}$, 2-15$^{(5)}$ \\
Pluto            & 16-26 $^{(8,9)}$ \\
Charon           &  9-14 $^{(8,9)}$ \\
\noalign{\smallskip}
\hline
\noalign{\smallskip}
Galilean satellites & 50-70 $^{(10)}$ \\
Saturnian satellites & 5-20 $^{(10)}$ \\
\noalign{\smallskip}
\hline
\noalign{\smallskip}
\multicolumn{2}{l}{\footnotesize (1) Fornasier et al.\ 2013; (2) Lellouch et al.\ 2017;}\\
\multicolumn{2}{l}{\footnotesize (3) Schindler et al.\ 2017; (4) Santos-Sanz et al.\ 2017;}\\
\multicolumn{2}{l}{\footnotesize (5) M\"uller et al.\ 2018; (6) P\'al et al.\ 2015;}\\
\multicolumn{2}{l}{\footnotesize (7) Kiss et al.\ 2018; (8,9) Lellouch et al.\ 2011/2016;}\\
\multicolumn{2}{l}{\footnotesize (10) Howett et al.\ 2010.}
\end{tabular}
\end{table} 
\end{center} 

\subsection{Emissivity}

A factor impacting surface/sub-surface temperatures calculated either by NEATM
or TPM is the so-called bolometric emissivity ($\epsilon_b$).
$\epsilon_b$ is defined as the Planck-function weighted average of the spectral emissivity $\epsilon(\lambda)$, where the latter is related to the spectral 
directional hemispheric reflectivity\footnote{In asteroid science, $\epsilon_b$ is routinely taken as 0.9 on the basis
that it is a typical value for silicates, given their mid-IR albedo. As extensively discussed by Myhrvold (2018), this assumption and other approximations
of the NEATM are questionable, but we don't dwell into this here.}. In addition, for a given surface/subsurface temperature profile, 
the radiation locally emitted at some wavelength depends on $\epsilon(\lambda)$. Many studies on asteroids have shown evidence for subdued 
fluxes at long wavelengths (submm/mm) compared to model expectations. This behaviour is  
usually termed ``emissivity effect", but quantitative estimates of this ``emissivity" depend on how much physics is put in the reference model. 
For example, the increasing transparency of ices towards long wavelengths (see e.g. Mishima et al. 1983 for H$_2$O ice) implies that
radiation progressively originates from the sub-surface. As dayside temperatures decrease with depth, this could lead
one to misinterpret an apparent decline of the T$_B$ with $\lambda$ in terms of a lower ``physical" emissivity. The mixing of horizontally
variable surface temperatures has the same effect of decreasing T$_B$ with $\lambda$ (until the Rayleigh-Jeans limit is hit).
Nonetheless, spectrally declining emissivities may also have physical causes. These include:
(i) reflection of upward-propagating thermal radiation at the surface (often described by Fresnel reflection for a dielectric surface);
(ii) particle scattering, which produces an emissivity minimum for particle sizes $a$ comparable to $\lambda$/4$\pi$;
(iii) volume scattering, a process in which subsurface inhomogeneities or voids on scales comparable or larger than the wavelength
     in weakly-absorbing medium cause multiple internal reflections
(iv) thermal inertia increase with depth over the first $\sim$cm into the surface, as observed for the Moon regolith (Keihm 1984),
     Mercury (Mitchell \& de Pater 1994) or asteroid 21 Lutetia (Gulkis et al.\ 2012).

These effects are still poorly characterized for TNOs. For Pluto/Charon, system-averaged T$_B$ decline from $\sim$ 53 K at 20 $\mu$m to  $\sim$ 35 K at 500 $\mu$m (Lellouch et al.\ 2016) and level out at this value across the mm/cm range (Butler et al. 2015, and in prep.).
Lellouch et al.\ (2016) showed that 35 K is lower than any expected temperature for the dayside surface/subsurface of Pluto and Charon, and inferred a ``true" spectral emissivity
decreasing steadily from 1 at 20-25 $\mu$m to $\sim$0.7 at 500 $\mu$m. This kind of behavior is usually not observed in asteroids
(when proper allowance is made for subsurface sounding, see Keihm et al. 2013), but is found in several outer solar system icy surfaces 
(e.g.\ Muhleman \& Berge 1991; Ostro et al.\ 2006; Janssen et al.\ 2009; Le Gall et al.\ 2014; Ries \& Janssen 2015) and in various kinds of
ice and snow on Earth (Hewison \& English 1999). Lellouch et al.\ (2016) concluded that the combination of a high dielectric constant ($\epsilon_r$ = 3-5) and a considerable surface material transparency (typical penetration depth $\sim$ 1 cm at 500 $\mu$m) was responsible for the effect.

Similar studies, albeit less detailed, have been performed on a dozen other TNO or Centaurs. This includes (i) Herschel/SPIRE 250-500 $\mu$m 
measurements of 9 objects strongly detected with Herschel/PACS 
(Fornasier et al.\ 2013) (ii) ALMA observations of 4 objects at 0.87 and 1.30 mm (Brown \& Butler 2017) and of six other at 1.30 mm only (Lellouch et al.\ 2017),
complemented by ancient and less accurate but useful data from IRAM and JCMT.
The SPIRE measurements generally did not detect the objects at 500 $\mu$m (and sometimes not 
at 350 or even 250 $\mu$m), implying an emissivity decline at wavelengths $>$200 $\mu$m, but the derived emissivity curves are very uncertain.
The much more accurate ALMA data show long-wavelength emissivity effects for all objects except Makemake, and lead to a mean {\em relative}  radio emissivity  $\epsilon_r$ (i.e. ratioed to the bolometric emissivity) of $\epsilon_r$ = 0.70$\pm$0.13. Lellouch et al.\ (2017) demonstrated the importance of including 
proper shape models and accounting for variable spin orientation to infer these emissivities. No correlation
was found between the radio emissivity and other (semi)-physical parameters such as diameter, color, composition, beaming factor, albedo, subsolar temperature, although
a possible trend of increasing emissivity with grain size was suggested.
A surprising result from Brown \& Butler (2017) is the sometimes marked (but not in consistent directions) spectral variation of emissivity for 
some objects over 0.87--1.3 mm. This is unexpected because in the framework of scattering effects, the emissivity minimum at $\lambda$ $\sim$ 4$\pi$$a$ 
is expected to be broad due to particle size distribution.

\section{Outlook}
\label{sec:outlook}

The Spitzer, WISE and Herschel missions provided a wealth of thermal data which was the key for 
the determination of thermophysical (sample-)properties of Centaurs and TNOs. But with the end of
the Herschel mission in 2013 it is not possible right now to detect these objects close to their thermal
emission peak in the far-IR regime. Individual objects can in principle be measured at mid-IR
(e.g., with SOFIA, possibly also from ground), but only the brightest ones are detectable
and with very little scientific gain over the existing observations from Spitzer and Herschel.
In the submm/mm wavelength range the situation is more comfortable (e.g.\ with ALMA, IRAM), but
measurements of a significant sample remain very time consuming. In addition, the unknowns about
the objects' emissivity make the interpretation more challenging. The next big step in thermal
emission observations of TNOs is expected to come with JWST\footnote{James Webb Space
Telescope: \url{https://jwst.stsci.edu}}. Norwood et al.\ (2016) stated that
MIRI\footnote{Mid-Infrared Instrument: \url{https://jwst.stsci.edu/instrumentation/miri}}
will be capable to characterize the thermal emission of TNOs in the 
18-25\,$\mu$m filters. Parker et al.\ (2016a) expect that the MIRI measurements will resolve
ambiguities in the thermophysical model studies and constrain thermal inertia and roughness
of TNO surfaces. JWST will also see the short-wavelengths thermal emission excess related to
hotter terrains (low-albedo regions) or originating from dark satellites.
In combination with existing Spitzer and Herschel measurements, the JWST thermal emission
measurements will lead to a significant refinement of the radiometric size-albedo solutions
and possibly provide evidence for seasonal volatile transport (e.g.\ on Sedna).
A major contribution to the thermal studies of Centaurs and TNOs is expected to come with
the Space Infrared Telescope for Cosmology and Astrophysics (SPICA). It will cover the
far-IR regime, but with much better sensitivity compared to Spitzer or Herschel. However,
the launch of this JAXA/ESA-proposed mission would not be before 2032.

\section*{Acknowledgement}

The research leading to these results has received funding from the European
Union's Horizon 2020 Research and Innovation Programme, under Grant Agreement no.\ 687378.

\newpage 

\clearpage 
{\scriptsize
 \begin{center}
{\renewcommand{\arraystretch}{1.4}
\begin{longtable}{|l|l|c|c|c|c|c|}
\caption[]{\noindent Absolute magnitude used in the modelling and derived radiometric diameters, geometric albedos, and beaming factor for TNOs and Centaurs.
The individual fluxes and the criteria for the dynamical classification for the objets are available in the TNOs are Cool public database:
\url{http://public-tnosarecool.lesia.obspm.fr/Published-observations.html}.
"med/out res" refers to resonants beside plutinos, not including Haumea and family members.
Geometric albedo and Abolute magnitude are given in the V filter, except for the 2 objects 2003 FE128 and 2007 RW10;$^*$: binary systems; $^{\P}$: absolute magnitude and albedo are in the R band; $^{\S}$: fixed $\eta$;  References: a) Duffard et al.\ 2014; b) Pal et al.\ 2015; c) Fornasier et al.\ 2013; d) Fornasier et al.\ 2014; e) Lacerda et al.\ 2014; f) Vilenius et al.\ 2014; g) Vilenius et al.\ 2012; h) Lellouch et al.\ 2013; i) Mommert,  2013 PhD thesis; j) Vilenius et al.\ 2018; k) Mommert et al.\ 2012; l) Kiss et al.\ 2018; m) M\"uller et al.\ 2010; n) Kiss et al.\ 2013; o) Santos-Sanz et al.\ 2012; p) Stansberry et al.\ 2008; q) Bauer et al.\ 2013; r) Nimmo et al.\ 2017; s) Buratti et al.\ 2015; t) Buratti et al.\ 2017; u) Stern et al.\ 2018; in cases where the uncertainty for H-mag is unknown, we assumed 0.3\,mag.}
\label{tbl:pv_all} \\
\hline
\multicolumn{1}{|l|}{\textbf{Object}} & \multicolumn{1}{l|}{\textbf{Dyn. class}} & \multicolumn{1}{c|}{\textbf{H$_{v}$}} & \multicolumn{1}{c|}{\textbf{Diam. (km)}} & \multicolumn{1}{c|}{\textbf{p$_{v}$}} & \multicolumn{1}{c|}{\textbf{$\eta$}} & \multicolumn{1}{c|}{\textbf{Ref.}}  \\ [0.5ex] \hline
\endfirsthead
\multicolumn{7}{c}{\tablename\ \thetable\ -- \textit{continued from previous page}} \\ \hline 
\multicolumn{1}{|l|}{\textbf{Object}} & \multicolumn{1}{l|}{\textbf{Dyn. class}} & \multicolumn{1}{c|}{\textbf{H$_{v}$}} & \multicolumn{1}{c|}{\textbf{Diam. (km)}} & \multicolumn{1}{c|}{\textbf{p$_{v}$}} & \multicolumn{1}{c|}{\textbf{$\eta$}} & \multicolumn{1}{|c|}{\textbf{Ref.}}  \\ \hline
\hline 
\endhead
\hline \multicolumn{4}{r}{{Continued on next page}} \\ 
\endfoot
\hline \hline
\endlastfoot
        2000 GM137 &        Centaur &14.36$\pm$0.38 &   8.6$^{+1.5}_{-1.5}$ & 0.043$^{+0.026}_{-0.016}$ & $^{\S}$1.20$^{+0.35}_{-0.35}$ &     a   \\ 
         2004 QQ26 &        Centaur & 9.53$\pm$0.36 &  79.0$^{+19.0}_{-19.0}$ & 0.044$^{+0.039}_{-0.014}$ & $^{\S}$1.20$^{+0.35}_{-0.35}$ &    a   \\
         2013 AZ60 &    Centaur/SDO &10.45$\pm$0.10 &  62.3$^{+5.3}_{-5.3}$ & 0.029$^{+0.006}_{-0.006}$ & 1.70$^{+0.90}_{-0.90}$ &        b   \\
  (2060)     Chiron &        Centaur & 5.92$\pm$0.20 & 215.6$^{+9.9}_{-9.9}$ & 0.167$^{+0.037}_{-0.030}$ & 0.95$^{+0.09}_{-0.10}$ &      c   \\
  (5145)     Pholus &        Centaur & 7.68$\pm$0.28 &  99.0$^{+15.0}_{-14.0}$ & 0.155$^{+0.076}_{-0.049}$ & 0.77$^{+0.16}_{-0.16}$ &     a   \\
  (7066)     Nessus &        Centaur & 9.51$\pm$0.22 &  57.0$^{+17.0}_{-14.0}$ & 0.086$^{+0.075}_{-0.034}$ & $^{\S}$1.20$^{+0.35}_{-0.35}$ &  a   \\
  (8405)    Asbolus &        Centaur & 9.13$\pm$0.25 &  85.0$^{+8.0}_{-9.0}$ & 0.056$^{+0.019}_{-0.015}$ & 0.97$^{+0.14}_{-0.18}$ &      a   \\
 (10199)   Chariklo &        Centaur & 7.40$\pm$0.25 & 238.0$^{+10.0}_{-10.0}$ & 0.042$^{+0.005}_{-0.005}$ & 1.12$^{+0.14}_{-0.14}$ &    c, d   \\
 (10370)   Hylonome &        Centaur & 9.51$\pm$0.08 &  74.0$^{+16.0}_{-16.0}$ & 0.051$^{+0.030}_{-0.017}$ & 1.29$^{+0.31}_{-0.31}$ &    a   \\
 (31824)     Elatus &        Centaur &10.40$\pm$0.09 &  49.8$^{+10.4}_{-9.8}$ & 0.049$^{+0.028}_{-0.016}$ & $^{\S}$1.20$^{+0.35}_{-0.35}$ &      a   \\
 (32532)    Thereus &        Centaur & 9.40$\pm$0.16 &  62.0$^{+3.0}_{-3.0}$ & 0.083$^{+0.016}_{-0.013}$ & 0.87$^{+0.08}_{-0.08}$ & a   \\
 (52872)    Okyrhoe &        Centaur &11.07$\pm$0.10 &  35.0$^{+3.0}_{-3.0}$ & 0.056$^{+0.012}_{-0.010}$ & 0.71$^{+0.12}_{-0.13}$ &        a   \\
 (52975)   Cyllarus &        Centaur & 9.02$\pm$0.15 &  56.0$^{+21.0}_{-18.0}$ & 0.139$^{+0.157}_{-0.064}$ & $^{\S}$1.20$^{+0.35}_{-0.35}$ &   a   \\
 (54598)     Bienor &        Centaur & 7.57$\pm$0.34 & 198.0$^{+6.0}_{-7.0}$ & 0.043$^{+0.016}_{-0.012}$ & 1.58$^{+0.07}_{-0.07}$ &       a   \\
 (55576)     Amycus &        Centaur & 8.27$\pm$0.11 & 104.0$^{+8.0}_{-8.0}$ & 0.083$^{+0.016}_{-0.015}$ & 1.00$^{+0.12}_{-0.13}$ &       a   \\
 (60558)   Echeclus &        Centaur & 9.78$\pm$0.14 &  64.6$^{+1.6}_{-1.6}$ & 0.052$^{+0.007}_{-0.007}$ & 0.86$^{+0.04}_{-0.04}$ &       a   \\
 (63252)  2001 BL41 &        Centaur &11.34$\pm$0.21 &  34.6$^{+6.6}_{-6.1}$ & 0.043$^{+0.028}_{-0.014}$ & $^{\S}$1.20$^{+0.35}_{-0.35}$ &    a   \\
 (83982)    Crantor &        Centaur & 9.03$\pm$0.16 &  59.0$^{+11.0}_{-12.0}$ & 0.121$^{+0.064}_{-0.038}$ & $^{\S}$1.20$^{+0.35}_{-0.35}$ &    a   \\
 (95626)  2002 GZ32 &        Centaur & 7.37$\pm$0.10 & 237.0$^{+8.0}_{-8.0}$ & 0.037$^{+0.004}_{-0.004}$ & 0.97$^{+0.05}_{-0.07}$ &      a   \\
(119315)  2001 SQ73 &        Centaur & 9.15$\pm$0.11 &  90.0$^{+23.0}_{-20.0}$ & 0.048$^{+0.030}_{-0.018}$ & $^{\S}$1.20$^{+0.35}_{-0.35}$ &    a   \\
(119976) 2002 VR130 &        Centaur &11.26$\pm$0.39 &  24.4$^{+5.4}_{-4.6}$ & 0.093$^{+0.066}_{-0.036}$ & $^{\S}$1.20$^{+0.35}_{-0.35}$ &   a   \\
(120061)   2003 CO1 &        Centaur & 9.07$\pm$0.05 &  94.0$^{+5.0}_{-5.0}$ & 0.049$^{+0.005}_{-0.006}$ & 1.23$^{+0.12}_{-0.11}$ &     a   \\
(136204)   2003 WL7 &        Centaur & 8.75$\pm$0.16 & 105.0$^{+6.0}_{-7.0}$ & 0.053$^{+0.010}_{-0.010}$ & 1.02$^{+0.07}_{-0.05}$ &     a   \\
(145486) 2005 UJ438 &        Centaur &11.14$\pm$0.32 &  16.0$^{+1.0}_{-2.0}$ & 0.256$^{+0.097}_{-0.076}$ & 0.34$^{+0.09}_{-0.08}$ &     a   \\
(248835) 2006 SX368 &        Centaur & 9.45$\pm$0.11 &  76.0$^{+2.0}_{-2.0}$ & 0.052$^{+0.007}_{-0.006}$ & 0.87$^{+0.04}_{-0.06}$ &     a   \\
(250112)  2002 KY14 &        Centaur &10.37$\pm$0.07 &  47.0$^{+3.0}_{-4.0}$ & 0.057$^{+0.011}_{-0.007}$ & $^{\S}$1.20$^{+0.35}_{-0.35}$ &  a   \\
(281371)  2008 FC76 &        Centaur & 9.44$\pm$0.10 &  68.0$^{+6.0}_{-7.0}$ & 0.067$^{+0.017}_{-0.011}$ & $^{\S}$1.20$^{+0.35}_{-0.35}$ &  a   \\
(447178)  2005 RO43 &        Centaur & 7.34$\pm$0.51 & 194.0$^{+10.0}_{-10.0}$ & 0.056$^{+0.036}_{-0.021}$ & 1.12$^{+0.05}_{-0.08}$ &     a   \\
        2000 CN105 &  Cold classical & 5.20$\pm$0.30 & 247.0$^{+63.0}_{-40.0}$ & 0.151$^{+0.070}_{-0.059}$ & $^{\S}$1.20$^{+0.35}_{-0.35}$ &      e    \\
        2001 QS322 &  Cold classical & 6.91$\pm$0.68 & 186.0$^{+99.0}_{-24.0}$ & 0.095$^{+0.531}_{-0.060}$ & $^{\S}$1.20$^{+0.35}_{-0.35}$ &   f   \\
        2001 RZ143 &  Cold classical & 6.69$\pm$0.13 & 140.0$^{+39.0}_{-33.0}$ & 0.191$^{+0.066}_{-0.045}$ & 0.75$^{+0.23}_{-0.19}$ &    g  \\
        2001 XR254 &  Cold classical & 6.05$\pm$0.15 & 221.0$^{+41.0}_{-71.0}$ & 0.136$^{+0.168}_{-0.044}$ & $^{\S}$1.20$^{+0.35}_{-0.35}$ &    f   \\
        2003 QR91$^*$ &  Cold classical & 6.55$\pm$0.56 & 280.0$^{+27.0}_{-30.0}$ & 0.054$^{+0.035}_{-0.028}$ & 1.20$^{+0.10}_{-0.12}$ &      f   \\
        2003 WU188 &  Cold classical & 5.96$\pm$0.64 & $<$ 220 & $> 0.15 $ & 1.20$^{+0.35}_{-0.35}$ &  f       \\
 (66652) Borasisi$^*$ &  Cold classical & 6.121$\pm$0.07 & 163.0$^{+32.0}_{-66.0}$ & 0.236$^{+0.438}_{-0.077}$ & 0.77$^{+0.19}_{-0.47}$ &      f   \\
(79360) Sila-Nunam &  Cold classical & 5.56$\pm$0.04 & 343.0$^{+42.0}_{-42.0}$ & 0.090$^{+0.027}_{-0.017}$ & 1.36$^{+0.21}_{-0.19}$ &    g   \\
 (88611) Teharonhiawako &  Cold classical & 6.00$\pm$0.13 & 220.0$^{+41.0}_{-44.0}$ & 0.145$^{+0.086}_{-0.045}$ & 1.08$^{+0.30}_{-0.28}$ &     f   \\
(119951)  2002 KX14 &  Cold classical & 4.86$\pm$0.10 & 455.0$^{+27.0}_{-27.0}$ & 0.097$^{+0.014}_{-0.013}$ & 1.79$^{+0.16}_{-0.15}$ &    g  \\
(120181) 2003 UR292 &  Cold classical & 7.40$\pm$0.40 & 136.0$^{+16.0}_{-26.0}$ & 0.105$^{+0.081}_{-0.033}$ & $^{\S}$1.20$^{+0.35}_{-0.35}$ &   f   \\
(135182) 2001 QT322 &  Cold classical & 7.29$\pm$0.67 & 159.0$^{+30.0}_{-47.0}$ & 0.085$^{+0.424}_{-0.052}$ & $^{\S}$1.20$^{+0.35}_{-0.35}$ &    f   \\
(275809) 2001 QY297 &  Cold classical & 5.86$\pm$0.31 & 229.0$^{+22.0}_{  -108.0}$ & 0.152$^{+0.439}_{-0.035}$ & 1.52$^{+0.22}_{-0.92}$ &   f  \\
(385266) 2001 QB298 &  Cold classical & 6.10$\pm$0.30 & 196.0$^{+71.0}_{-53.0}$ & 0.167$^{+0.162}_{-0.082}$ & $^{\S}$1.20$^{+0.35}_{-0.35}$ &    i    \\
(385437)  2003 GH55 &  Cold classical & 6.43$\pm$0.12 & 178.0$^{+21.0}_{-56.0}$ & 0.150$^{+0.182}_{-0.031}$ & $^{\S}$1.20$^{+0.35}_{-0.35}$ &   f   \\
(469438)  2002 GV31 &  Cold classical & 6.10$\pm$0.60 & $<$180.0  & $>$0.019 &  $^{\S}$1.20$^{+0.35}_{-0.35}$ &  f   \\
(469514) 2003 QA91$^*$ &  Cold classical & 5.76$\pm$0.63 & 260.0$^{+30.0}_{-36.0}$ & 0.130$^{+0.119}_{-0.075}$ & 0.83$^{+0.10}_{-0.15}$ & f   \\
(469705) 2005 EF298 &  Cold classical & 6.40$\pm$0.50 & 174.0$^{+27.0}_{-32.0}$ & 0.160$^{+0.130}_{-0.070}$ & $^{\S}$1.20$^{+0.35}_{-0.35}$ & g  \\
(508869) 2002 VT130 &  Cold classical & 5.80$\pm$0.30 & 324.0$^{+57.0}_{-68.0}$ & 0.097$^{+0.098}_{-0.049}$ & $^{\S}$1.20$^{+0.35}_{-0.35}$ &   i    \\
         1996 TS66 &  Hot classical & 6.50$\pm$0.05 & 159.0$^{+44.0}_{-46.0}$ & 0.179$^{+0.173}_{-0.070}$ & 0.75$^{+0.21}_{-0.27}$ &   f  \\
         2001 KA77 &  Hot classical & 5.64$\pm$0.12 & 310.0$^{170.0}_{-60.0}$ & 0.099$^{+0.052}_{-0.056}$ & 2.52$^{+0.18}_{-0.83}$ & g  \\
        2001 QC298 &  Hot classical & 6.26$\pm$0.32 & 303.0$^{+27.0}_{-30.0}$ & 0.061$^{+0.027}_{-0.017}$ & 0.99$^{+0.08}_{-0.10}$ &    f   \\
        2001 QD298 &  Hot classical & 6.71$\pm$0.17 & 233.0$^{+27.0}_{-63.0}$ & 0.067$^{+0.062}_{-0.014}$ & $^{\S}$1.26$^{+0.35}_{-0.35}$ & g   \\
         2002 GH32 &  Hot classical & 6.58$\pm$0.28 & $<$180 & $>$ 0.13 &   $^{\S}$1.26$^{+0.35}_{-0.35}$ &  f  \\
 (19308)  1996 TO66 &  Hot classical & 4.81$\pm$0.14 & $<$ 330.0 & $>$0.200 & $^{\S}$1.74$^{+0.17}_{-0.17}$ &     j   \\
 (19521)      Chaos &  Hot classical & 5.00$\pm$0.06 & 600.0$^{140.0}_{  -130.0}$ & 0.050$^{+0.030}_{-0.016}$ & 2.20$^{+1.20}_{-1.10}$ &    g  \\
 (20000)     Varuna &  Hot classical & 3.76$\pm$0.035&668.0$^{154.0}_{-86.0}$ & 0.127$^{+0.040}_{-0.042}$ & 2.18$^{+1.04}_{-0.49}$ &   h   \\
 (24835)  1995 SM55 &  Hot classical & 4.49$\pm$0.035& $<$280.0  & $>$0.360 & $^{\S}$1.74$^{+0.17}_{-0.17}$ &     j    \\
 (35671) 1998 SN165 &  Hot classical & 5.707$\pm$0.085&393.0$^{+39.0}_{-38.0}$ & 0.060$^{+0.019}_{-0.013}$ & $^{\S}$1.23$^{+0.35}_{-0.35}$ & g    \\
 (50000) Quaoar$^*$ &  Hot classical & 2.73$\pm$0.06 & 1073.6$^{+37.9}_{-37.9}$ & 0.127$^{+0.010}_{-0.009}$ & 1.73$^{+0.08}_{-0.08}$ &     c    \\
 (55565) 2002 AW197 &  Hot classical & 3.57$\pm$0.05 &768.0$^{+39.0}_{-38.0}$ & 0.112$^{+0.012}_{-0.011}$ & 1.29$^{+0.13}_{-0.10}$ &     f   \\
 (55636) 2002 TX300 &  Hot classical & 3.37$\pm$0.05 & 323.0$^{+95.0}_{-37.0}$ & 0.760$^{+0.180}_{-0.450}$ & 1.80$^{+0.50}_{-0.90}$ &     j   \\
 (55637) 2002 UX25$^*$ &  Hot classical & 3.87$\pm$0.02 & 697.2$^{+23.0}_{-24.5}$ & 0.107$^{+0.008}_{-0.008}$ & 1.07$^{+0.05}_{-0.05}$ & h \\
 (78799)  2002 XW93 &  Hot classical & 5.40$\pm$0.70 & 565.0$^{+71.0}_{-73.0}$ & 0.038$^{+0.043}_{-0.025}$ & 0.79$^{+0.27}_{-0.24}$ &   g  \\
 (86177) 1999 RY215 &  Hot classical & 7.235$\pm$0.093&263.0$^{+29.0}_{-37.0}$ & 0.039$^{+0.012}_{-0.007}$ & $^{\S}$1.20$^{+0.35}_{-0.35}$ & g    \\
 (90568)   2004 GV9 &  Hot classical & 4.23$\pm$0.10 & 680.0$^{+34.0}_{-34.0}$ & 0.077$^{+0.008}_{-0.008}$ & 1.93$^{+0.09}_{-0.07}$ &   g  \\
(120178)  2003 OP32 &  Hot classical & 4.10$\pm$0.07 & 274.0$^{+47.0}_{-25.0}$ & 0.540$^{+0.110}_{-0.150}$ & $^{\S}$1.74$^{+0.17}_{-0.17}$ & j   \\
(120347) Salacia$^*$ &  Hot classical & 4.25$\pm$0.05 & 901.0$^{+45.0}_{-45.0}$ & 0.044$^{+0.004}_{-0.004}$ & 1.16$^{+0.03}_{-0.03}$ &   l    \\
(136108) Haumea$^*$ &  Hot classical & 0.43$\pm$0.11 & 1239.5$^{+68.7}_{-57.8}$ & 0.804$^{+0.062}_{-0.095}$ & 0.95$^{+0.33}_{-0.26}$ &  c  \\
(136472)   Makemake &  Hot classical & 0.14$\pm$0.05 & 1430.0$^{+ 9.0}_{-9.0}$ & 0.770$^{+0.020}_{-0.020}$ & 2.29$^{+0.46}_{-0.40}$ &       h   \\
(138537)  2000 OK67 &  Hot classical & 6.47$\pm$0.13 & 164.0$^{+33.0}_{-45.0}$ & 0.169$^{+0.159}_{-0.052}$ & $^{\S}$1.20$^{+0.35}_{-0.35}$ & f   \\
(145452)  2005 RN43 &  Hot classical & 3.89$\pm$0.05 & 679.0$^{+55.0}_{-73.0}$ & 0.107$^{+0.029}_{-0.018}$ & $^{\S}$1.20$^{+0.35}_{-0.35}$ & g  \\
(145453)  2005 RR43 &  Hot classical & 4.13$\pm$0.08 & 300.0$^{+43.0}_{-34.0}$ & 0.440$^{+0.120}_{-0.100}$ & $^{\S}$1.74$^{+0.17}_{-0.17}$ &  j   \\
(148780) Altjira$^*$ &  Hot classical & 6.44$\pm$0.14 & 331.0$^{+51.0}_{  -187.0}$ & 0.043$^{+0.183}_{-0.009}$ & 1.62$^{+0.24}_{-0.83}$ &  f   \\
(174567)  Varda$^*$ &  Hot classical & 3.61$\pm$0.05 & 792.0$^{+91.0}_{-84.0}$ & 0.102$^{+0.024}_{-0.020}$ & 0.84$^{+0.28}_{-0.22}$ &   f   \\
(182934)  2002 GJ32 &  Hot classical & 6.16$\pm$0.13 & 416.0$^{+81.0}_{-78.0}$ & 0.035$^{+0.019}_{-0.011}$ & 2.05$^{+0.38}_{-0.36}$ &    f   \\
(202421) 2005 UQ513 &  Hot classical & 3.87$\pm$0.14 & 498.0$^{+63.0}_{-75.0}$ & 0.202$^{+0.084}_{-0.049}$ & $^{\S}$1.27$^{+0.35}_{-0.35}$ & g  \\
(230965) 2004 XA192 &  Hot classical & 4.42$\pm$0.63 & 339.0$^{120.0}_{-95.0}$ & 0.260$^{+0.340}_{-0.150}$ & 0.62$^{+0.79}_{-0.49}$ & f   \\
(307251)  2002 KW14 &  Hot classical & 5.88$\pm$0.11 & 161.0$^{+35.0}_{-40.0}$ & 0.310$^{+0.281}_{-0.094}$ & $^{\S}$1.20$^{+0.35}_{-0.35}$ &  f   \\
(307261)   2002 MS4 &  Hot classical & 4.00$\pm$0.60 & 934.0$^{+47.0}_{-47.0}$ & 0.051$^{+0.036}_{-0.022}$ & 1.06$^{+0.06}_{-0.06}$ &  g  \\
(307616)  2003 QW90 &  Hot classical & 5.00$\pm$0.30 & 401.0$^{+63.0}_{-48.0}$ & 0.084$^{+0.026}_{-0.022}$ & $^{\S}$1.11 &       e   \\
(416400) 2003 UZ117 &  Hot classical & 5.23$\pm$0.15 & 222.0$^{+57.0}_{-42.0}$ & 0.290$^{+0.160}_{-0.110}$ & $^{\S}$1.74$^{+0.17}_{-0.17}$ & j   \\
(444030)  2004 NT33 &  Hot classical & 4.74$\pm$0.11 & 423.0$^{+87.0}_{-80.0}$ & 0.125$^{+0.069}_{-0.039}$ & 0.69$^{+0.460}_{-0.32}$ &            f   \\
(469306) 1999 CD158 &  Hot classical & 5.35$\pm$0.67 & $<$310.0 & $>$0.130  & $^{\S}$1.20$^{+0.35}_{-0.35}$ &    j   \\
(469615) 2004 PT107 &  Hot classical & 6.33$\pm$0.11 & 400.0$^{+45.0}_{-51.0}$ & 0.032$^{+0.011}_{-0.007}$ & $^{\S}$1.53$^{+0.35}_{-0.35}$ &   f   \\
         2001 KD77 &        Plutino & 6.42$\pm$0.08 & 232.3$^{+40.5}_{-39.4}$ & 0.089$^{+0.044}_{-0.027}$ & $^{\S}$1.20$^{+0.35}_{-0.35}$ &   k    \\
         2002 XV93 &        Plutino & 5.42$\pm$0.46 & 549.2$^{+21.7}_{-23.0}$ & 0.040$^{+0.020}_{-0.015}$ & 1.24$^{+0.06}_{-0.06}$ &     k    \\
        2003 UT292 &        Plutino & 6.85$\pm$0.68 & 185.6$^{+17.9}_{-18.0}$ & 0.067$^{+0.068}_{-0.034}$ & $^{\S}$1.20$^{+0.35}_{-0.35}$ &   k    \\
 (15820)    1994 TB &        Plutino & 7.934$\pm$0.354& 85.0$^{+36.0}_{-28.0}$ & 0.172$^{+0.258}_{-0.097}$ & 1.260$^{+0.97}_{-0.65}$ & h    \\
 (15875)  1996 TP66 &        Plutino & 7.51$\pm$0.09 & 154.0$^{+28.8}_{-33.7}$ & 0.074$^{+0.063}_{-0.031}$ & $^{\S}$1.20$^{+0.35}_{-0.35}$ &  k    \\
 (28978)      Ixion &        Plutino & 3.83$\pm$0.04 & 617.0$^{+19.0}_{-20.0}$ & 0.141$^{+0.011}_{-0.011}$ & 0.91$^{+0.04}_{-0.06}$ & h    \\
 (33340)  1998 VG44 &        Plutino & 6.67$\pm$0.04 & 248.0$^{+43.0}_{-41.0}$ & 0.063$^{+0.026}_{-0.017}$ & 1.55$^{+0.58}_{-0.38}$ & h    \\
 (38628)   Huya$^*$ &        Plutino & 5.04$\pm$0.03 & 458.0$^{+ 9.2}_{-9.2}$ & 0.083$^{+0.004}_{-0.004}$ & 0.93$^{+0.02}_{-0.02}$ &     c   \\
 (47171) 1999 TC36$^*$ &        Plutino & 5.41$\pm$0.10 & 393.1$^{+25.2}_{-26.8}$ & 0.079$^{+0.013}_{-0.011}$ & 1.10$^{+0.07}_{-0.08}$ &     k    \\
 (47932) 2000 GN171 &        Plutino & 6.45$\pm$0.34 & 147.1$^{+20.7}_{-17.8}$ & 0.215$^{+0.093}_{-0.070}$ & 1.11$^{+0.24}_{-0.21}$ &     k    \\
 (55638)  2002 VE95 &        Plutino & 5.70$\pm$0.06 & 249.8$^{+13.5}_{-13.1}$ & 0.149$^{+0.019}_{-0.016}$ & 1.40$^{+0.12}_{-0.11}$ &     k    \\
 (84719) 2002 VR128 &        Plutino & 5.58$\pm$0.37 & 448.5$^{+42.1}_{-43.2}$ & 0.052$^{+0.027}_{-0.018}$ & $^{\S}$1.20$^{+0.35}_{-0.35}$ &  k    \\
 (84922)   2003 VS2 &        Plutino & 4.11$\pm$0.38 & 523.0$^{+35.1}_{-34.4}$ & 0.147$^{+0.063}_{-0.043}$ & 1.57$^{+0.30}_{-0.23}$ &     k    \\
 (90482)  Orcus$^*$ &        Plutino & 2.31$\pm$0.03 & 958.4$^{+22.9}_{-22.9}$ & 0.231$^{+0.018}_{-0.011}$ & 0.97$^{+0.05}_{-0.02}$ &     c   \\
(120216)  2004 EW95 &        Plutino & 6.69$\pm$0.35 & 291.1$^{+20.3}_{-25.9}$ & 0.044$^{+0.021}_{-0.015}$ & $^{\S}$1.20$^{+0.35}_{-0.35}$ &  k    \\
(120348) 2004 TY364 &        Plutino & 4.52$\pm$0.07 & 512.0$^{+37.0}_{-40.0}$ & 0.107$^{+0.020}_{-0.015}$ & 1.55$^{+0.15}_{-0.10}$ & h    \\
(133067) 2003 FB128 &        Plutino & 6.80$\pm$0.30 & 186.0$^{+27.0}_{-29.0}$ & 0.074$^{+0.035}_{-0.021}$ & $^{\S}$1.20$^{+0.35}_{-0.35}$ &  e    \\
(144897)  2004 UX10 &        Plutino & 4.75$\pm$0.16 & 398.1$^{+32.6}_{-39.3}$ & 0.141$^{+0.044}_{-0.031}$ & $^{\S}$1.20$^{+0.35}_{-0.35}$ &  k    \\
(175113) 2004 PF115 &        Plutino & 4.54$\pm$0.25 & 468.2$^{+38.6}_{-49.1}$ & 0.123$^{+0.043}_{-0.033}$ & $^{\S}$1.20$^{+0.35}_{-0.35}$ &  k    \\
(208996) 2003 AZ84$^*$ &        Plutino & 3.74$\pm$0.08 & 727.0$^{+61.9}_{-66.5}$ & 0.107$^{+0.023}_{-0.016}$ & 1.05$^{+0.19}_{-0.15}$ &     k    \\
(307463) 2002 VU130 &        Plutino & 5.47$\pm$0.83 & 252.9$^{+33.6}_{-31.3}$ & 0.179$^{+0.202}_{-0.103}$ & $^{\S}$1.20$^{+0.35}_{-0.35}$ &  k    \\
(450265) 2003 WU172 &        Plutino & 6.30$\pm$0.30 & 312.0$^{+ 0.0}_{ 0.0}$ & 0.039$^{+0.000}_{ 0.000}$ &  $^{\S}$2.60$^{+0.35}_{-0.35}$ &  e    \\
(455502) 2003 UZ413 &        Plutino & 5.70$\pm$0.30 & 670.0$^{+84.0}_{-82.0}$ & 0.070$^{+0.022}_{-0.015}$ & $^{\S}$1.20$^{+0.35}_{-0.35}$ &  e    \\
(469372) 2001 QF298 &        Plutino & 5.43$\pm$0.07 & 408.2$^{+40.2}_{-44.9}$ & 0.071$^{+0.020}_{-0.014}$ & $^{\S}$1.20$^{+0.35}_{-0.35}$ &  k    \\
(469987) 2006 HJ123 &        Plutino & 5.32$\pm$0.66 & 216.4$^{+29.7}_{-34.2}$ & 0.281$^{+0.259}_{-0.152}$ & $^{\S}$1.20$^{+0.35}_{-0.35}$ &  k    \\
(126154) 2001 YH140 &        med res & 5.80$\pm$0.20 & 349.0$^{+81.0}_{-81.0}$ & 0.080$^{+0.050}_{-0.050}$ & $^{\S}$1.20$^{+0.35}_{-0.35}$ &  m    \\
         2002 GP32 &        out res & 6.90$\pm$0.30 & 201.0$^{+25.0}_{-29.0}$ & 0.091$^{+0.061}_{-0.024}$ & $^{\S}$1.20$^{+0.35}_{-0.35}$ &    e    \\
 (26308) 1998 SM165 &        out res & 6.02$\pm$0.08&291.0$^{+22.0}_{-26.0}$ & 0.083$^{+0.018}_{-0.013}$ & 1.55$^{+0.11}_{-0.10}$ & h    \\
 (26375)   1999 DE9 &        out res & 5.16$\pm$0.05 &311.0$^{+29.0}_{-32.0}$ & 0.163$^{+0.041}_{-0.026}$ & 0.71$^{+0.10}_{-0.11}$ & h    \\
 (42301) 2001 UR163 &        out res & 4.10$\pm$0.30 & 352.0$^{+85.0}_{-53.0}$ & 0.209$^{+0.082}_{-0.074}$ & $^{\S}$1.20$^{+0.35}_{-0.35}$ &    e    \\
 (82075) 2000 YW134 &        out res & 4.88$\pm$0.05 & $<$500 & $>$0.08 &  $^{\S}$1.20$^{+0.35}_{-0.35}$ & m \\
 (84522) 2002 TC302 &        out res & 4.17$\pm$0.10 & 584.1$^{+105.6}_{-88.0}$ & 0.115$^{+0.047}_{-0.033}$ & 1.09$^{+0.37}_{-0.25}$ &   c     \\
(119979) 2002 WC19$^*$ &        out res & 4.88$\pm$0.07 & 348.0$^{+45.0}_{-45.0}$ & 0.167$^{+0.052}_{-0.037}$ & 1.12$^{+0.16}_{-0.17}$ & h    \\
(143707) 2003 UY117 &        out res & 5.70$\pm$0.30 & 247.0$^{+30.0}_{-29.0}$ & 0.126$^{+0.039}_{-0.028}$ & $^{\S}$1.20$^{+0.35}_{-0.35}$ &   e    \\
(225088) 2007 OR10$^*$ &     out res & 2.34$\pm$0.05 & 1252.0$^{43.0}_{-42.0}$ & 0.13$^{+0.01}_{-0.01}$ & 1.20$^{+0.35}_{-0.35}$ &   l    \\
(308379)  2005 RS43 &        out res & 5.00$\pm$0.30 & 271.0$^{+45.0}_{-40.0}$ & 0.193$^{+0.071}_{-0.053}$ & $^{\S}$1.20$^{+0.35}_{-0.35}$ &     e    \\
(469505) 2003 FE128 &        out res & $^{\P}$6.30$\pm$0.30 & 157.0$^{+60.0}_{-7.0}$ & $^{\P}$0.167$^{+0.085}_{-0.072}$ & $^{\S}$1.20$^{+0.35}_{-0.35}$ & f    \\
(471143) 2010 EK139 &        out res & 3.80$\pm$0.10 & 433.0$^{+63.0}_{-64.0}$ & 0.297$^{+0.113}_{-0.078}$ & 0.60$^{+0.33}_{-0.25}$ & h    \\
         2012 DR30 &            SDO & 7.04$\pm$0.35 & 188.0$^{+ 9.4}_{-9.4}$ & 0.076$^{+0.031}_{-0.025}$ & 0.81$^{+0.074}_{-0.062}$ &   n    \\
 (15874)  1996 TL66 &            SDO & 5.39$\pm$0.12 & 339.0$^{+20.0}_{-20.0}$ & 0.110$^{+0.021}_{-0.015}$ & 1.15$^{+0.08}_{-0.05}$ &   o    \\
 (26181)  1996 GQ21 &            SDO & 5.20$\pm$0.30 & 349.0$^{+43.0}_{-49.0}$ & 0.127$^{+0.043}_{-0.026}$ & $^{\S}$1.20$^{+0.35}_{-0.35}$ &  i    \\
 (29981)  1999 TD10 &            SDO & 8.93$\pm$0.30 & 103.7$^{+13.6}_{-13.5}$ & 0.044$^{+0.014}_{-0.010}$ & 1.64$^{+0.32}_{ 0.31}$ &    p   \\
 (42355) Typhon$^*$ &            SDO & 7.72$\pm$0.04 & 185.0$^{+ 7.0}_{-7.0}$ & 0.044$^{+0.003}_{-0.003}$ & 1.48$^{+0.07}_{-0.07}$ &    o    \\
 (44594)   1999 OX3 &            SDO & 7.72$\pm$0.09 & 135.0$^{+13.0}_{-12.0}$ & 0.081$^{+0.018}_{-0.015}$ & 1.04$^{+0.27}_{-0.22}$ & h    \\
 (48639) 1995 TL8$^*$ &          SDO & 5.29$\pm$0.06 & 244.0$^{+82.0}_{-63.0}$ & 0.231$^{+0.189}_{-0.102}$ & 1.38$^{+0.80}_{-0.49}$ & h    \\
 (65489)   Ceto$^*$ &            SDO & 6.54$\pm$0.06 & 281.0$^{+11.0}_{-11.0}$ & 0.056$^{+0.006}_{-0.006}$ & 1.04$^{+0.05}_{-0.05}$ &      o    \\
 (73480)  2002 PN34 &            SDO & 8.66$\pm$0.03 & 112.0$^{+ 7.0}_{-7.0}$ & 0.049$^{+0.006}_{-0.006}$ & 1.02$^{+0.07}_{-0.09}$ &    o    \\
 (82158) 2001 FP185 &            SDO & 6.39$\pm$0.07 & 332.0$^{+31.0}_{-24.0}$ & 0.046$^{+0.007}_{-0.007}$ & 1.23$^{+0.24}_{-0.19}$ &    o    \\
(127546)  2002 XU93 &            SDO & 8.11$\pm$0.10 & 164.0$^{+ 9.0}_{-9.0}$ & 0.038$^{+0.004}_{-0.004}$ & 1.12$^{+0.05}_{-0.08}$ &    o    \\
(309239)  2007 RW10 &            SDO & $^{\P}$6.39$\pm$0.61 & 247.0$^{+30.0}_{-30.0}$ & $^{\P}$0.083$^{+0.068}_{-0.039}$ & $^{\S}$1.20$^{+0.35}_{-0.35}$ &   o    \\
 (40314)  1999 KR16 &       Detached & 6.24$\pm$0.15 & 232.0$^{+34.0}_{-36.0}$ & 0.105$^{+0.049}_{-0.027}$ & $^{\S}$1.20$^{+0.35}_{-0.35}$  & j   \\
 (90377)      Sedna &       Detached & 1.83$\pm$0.05 & 906.0$^{314.0}_{-258.0}$ & 0.410$^{+0.390}_{-0.190}$ & 0.72$^{+0.78}_{-0.54}$ & h    \\
(120132) 2003 FY128 &       Detached & 5.09$\pm$0.09 & 460.0$^{+21.0}_{-21.0}$ & 0.079$^{+0.010}_{-0.010}$ & 1.07$^{+0.08}_{-0.08}$ &    o    \\
(136199)   Eris$^*$ &       Detached &-1.12$\pm$0.03 & 2326.0$^{+12.0}_{-12.0}$ & 0.960$^{+0.040}_{-0.040}$ & 0.87$^{+0.26}_{-0.41}$ & h    \\
(145480) 2005 TB190 &       Detached & 4.40$\pm$0.11 & 464.0$^{+62.0}_{-62.0}$ & 0.148$^{+0.051}_{-0.036}$ & $^{\S}$1.20$^{+0.35}_{-0.35}$ &  o    \\
(229762) 2007 UK126 &       Detached & 3.69$\pm$0.10 & 599.0$^{+77.0}_{-77.0}$ & 0.167$^{+0.058}_{-0.038}$ & $^{\S}$1.20$^{+0.35}_{-0.35}$ &    o    \\
(303775) 2005 QU182 &       Detached & 3.80$\pm$0.32 & 416.0$^{+73.0}_{-73.0}$ & 0.328$^{+0.160}_{-0.109}$ & $^{\S}$1.20$^{+0.35}_{-0.35}$ &    o    \\
(470316)  2007 OC10 &       Detached & 5.43$\pm$0.10 & 309.0$^{+37.0}_{-37.0}$ & 0.127$^{+0.040}_{-0.028}$ & $^{\S}$1.20$^{+0.35}_{-0.35}$ &    o    \\
%--------------
(330759) 2008 SO218      & Centaur & 12.8$\pm$0.3 &   11.8  $\pm$   0.4   &  0.097 $\pm$  0.017 &  0.823 $\pm$  0.046 & q \\ % Free	       X0759	
  2008 JS14              & Centaur & 13.2$\pm$0.3 &   14.5  $\pm$   1.8   &  0.044 $\pm$  0.019 &  1.046 $\pm$  0.186 & q \\ % Free	       K08J14S  
  2010 CR140             & Centaur & 15.5$\pm$0.3 &    7.5  $\pm$   1.4   &  0.020 $\pm$  0.01  &  1.111 $\pm$  0.283 & q \\ % Free	       K10CE0R  
  2010 HU20              & Centaur & 13.0$\pm$0.3 &   10.513$\pm$   1.1   &  0.101 $\pm$  0.024 &  0.976 $\pm$  0.162 & q \\ % Free	       K10H20U  
  2010 LG61              & Centaur & 18.5$\pm$0.3 &    0.89 $\pm$   0.19  &  0.089 $\pm$  0.056 &  1.00  $\pm$  0.400 & q \\ % Free	       K10L61G  
  2010 OR1               & Centaur & 16.2$\pm$0.3 &    3.25 $\pm$   0.64  &  0.055 $\pm$  0.013 &  0.831 $\pm$  0.146 & q \\ % Free	       K10O01R  
  2010 OM101             & Centaur & 17. $\pm$0.3 &    3.12 $\pm$   0.17  &  0.029 $\pm$  0.005 &  1.054 $\pm$  0.105 & q \\ % Free	       K10OA1M  
  2010 PO58              & Centaur & 14.5$\pm$0.3 &    8.88 $\pm$   0.63  &  0.035 $\pm$  0.007 &  0.915 $\pm$  0.093 & q \\ % Free	       K10P58O  
167P/CINEOS              & Centaur &  9.7$\pm$0.3 &   66.17 $\pm$  22.9   &  0.053 $\pm$  0.019 &  0.8   $\pm$  0.360 & q \\ % Fixed eta W4	    167P     
29P/Schwassm.-W.~1       & Centaur &  9. $\pm$0.3 &   46.   $\pm$  13.    &  0.033 $\pm$  0.015 &  0.64  $\pm$  0.29  & q \\ % Free		    29P      
(148975) 2001 XA255      & Centaur & 11.2$\pm$0.3 &   37.7  $\pm$  10.5   &  0.041 $\pm$  0.014 &  0.703 $\pm$  0.186 & q \\ % Free		    E8975    
(309139) 2006 XQ51       & Centaur &  9.8$\pm$0.3 &   39.1  $\pm$  15.7   &  0.139 $\pm$  0.058 &  0.8   $\pm$  0.456 & q \\ % Fixed eta W4	    U9139    
(310071) 2010 KR59       & Centaur &  7.7$\pm$0.3 &  110.06 $\pm$  30.82  &  0.121 $\pm$  0.037 &  0.8   $\pm$  0.324 & q \\ % Fixed eta W4	    V0071    
(309737) 2008 SJ236      & Centaur & 12.2$\pm$0.3 &   17.7  $\pm$   1.5   &  0.074 $\pm$  0.021 &  0.800 $\pm$  0.110 & q \\ % Free		    U9737    
(328884) 2010 LJ109      & Centaur & 10.1$\pm$0.3 &   44.2  $\pm$   3.8   &  0.083 $\pm$  0.021 &  0.748 $\pm$  0.103 & q \\ % Free		    W8884    
(332685) 2009 HH36       & Centaur & 10.6$\pm$0.3 &   33.0  $\pm$   2.8   &  0.078 $\pm$  0.018 &  0.739 $\pm$  0.095 & q \\ % Free		    X2685    
(342842) 2008 YB3        & Centaur &  9.5$\pm$0.3 &   67.1  $\pm$   1.0   &  0.062 $\pm$  0.012 &  0.839 $\pm$  0.012 & q \\ % Free		    Y2842    
(346889) 2009 QV38       & Centaur & 11.8$\pm$0.3 &   23.2  $\pm$   9.5   &  0.062 $\pm$  0.049 &  0.8   $\pm$  0.389 & q \\ % Fixed eta W3	    Y6889    
  2007 VH305             & Centaur & 11.6$\pm$0.3 &   23.8  $\pm$   8.0   &  0.070 $\pm$  0.036 &  0.8   $\pm$  0.384 & q \\ % Fixed eta W3, W4     K07VU5H  
  2008 HY21              & Centaur & 12.1$\pm$0.3 &   24.0  $\pm$   1.5   &  0.044 $\pm$  0.010 &  1.22  $\pm$  0.094 & q \\ % Free		    K08H21Y  
  2010 BL4               & Centaur & 11.9$\pm$0.3 &   15.7  $\pm$   3.2   &  0.114 $\pm$  0.052 &  0.8   $\pm$  0.333 & q \\ % Fixed eta W4	    K10B04L  
  2010 ES65              & Centaur & 11.8$\pm$0.3 &   26.9  $\pm$   7.9   &  0.049 $\pm$  0.024 &  0.8   $\pm$  0.28  & q \\ % Fixed eta W3, W4     K10E65S  
  2010 FH92              & Centaur & 11.7$\pm$0.3 &   28.0  $\pm$   0.6   &  0.047 $\pm$  0.007 &  0.730 $\pm$  0.023 & q \\ % Free		    K10F92H  
  2010 RM64              & Centaur & 11.0$\pm$0.3 &   21.0  $\pm$   2.0   &  0.159 $\pm$  0.048 &  0.85  $\pm$  0.144 & q \\ % Free		    K10R64M  
  2010 TH                & Centaur &  9.2$\pm$0.3 &   69.9  $\pm$  24.2   &  0.078 $\pm$  0.033 &  0.8   $\pm$  0.363 & q \\ % Fixed eta W4	    K10T00H  
  2011 MM4               & Centaur &  9.3$\pm$0.3 &   63.7  $\pm$   6.2   &  0.083 $\pm$  0.024 &  0.841 $\pm$  0.119 & q \\ % Free		    K11M04M  
  2005 VJ119             &     SDO & 10.6$\pm$0.3 &   28.5  $\pm$   6.9   &  0.126 $\pm$  0.060 &  0.8   $\pm$  0.30  & q \\ % Fixed eta W4	    K05VB9J  
  2010 BK118             &     SDO & 10.2$\pm$0.3 &   46.4  $\pm$   1.8   &  0.068 $\pm$  0.013 &  0.821 $\pm$  0.043 & q \\ % Free		    K10BB8K  
  2010 GW64              &     SDO & 14.9$\pm$0.3 &    6.42 $\pm$   0.38  &  0.047 $\pm$  0.012 &  0.795 $\pm$  0.075 & q \\ % Free		    K10G64W  
  2010 GW147             &     SDO & 13.2$\pm$0.3 &   15.9  $\pm$   0.7   &  0.037 $\pm$  0.006 &  0.869 $\pm$  0.056 & q \\ % Free		    K10GE7W  
  2010 JH124             &     SDO & 14.6$\pm$0.3 &    7.04 $\pm$   0.74  &  0.052 $\pm$  0.024 &  0.959 $\pm$  0.164 & q \\ % Free		    K10JC4H  
C/2010 KW7 (WISE)        &     SDO & 15.5$\pm$0.3 &    4.87 $\pm$   0.22  &  0.047 $\pm$  0.011 &  0.75  $\pm$  0.06  & q \\ % Free		    K10K07W  
  2010 WG9               &     SDO &  8.1$\pm$0.3 &  112.7  $\pm$  61.9   &  0.074 $\pm$  0.080 &  0.8   $\pm$  0.423 & q \\ % Fixed eta W4	    K10W09G  
(336756) 2010 NV1        &     SDO & 10.5$\pm$0.3 &   44.2  $\pm$   8.0   &  0.057 $\pm$  0.030 &  0.661 $\pm$  0.168 & q \\ % Free		    X6756    
C/2011 KP36 (Spacewatch) & Centaur &  9.4$\pm$0.3 &   55.1  $\pm$  19.4   &  0.101 $\pm$  0.062 &  0.8   $\pm$  0.4   & q \\ % Fixed eta W4  CK11K36P 
Pluto                    & Plutino & -0.7         & 2376.6  $\pm$   3.2   & \multicolumn{2}{l|}{0.52$\pm$0.03 (0.08-1.0)} & r,s,t \\
Charon                   & Plutino &  1.0         & 1212.0  $\pm$   1.0   & \multicolumn{2}{l|}{0.41$\pm$0.02 (0.11-0.73)} & t,u \\
\hline
\end{longtable}
}
\end{center}
}
\begin{list}{}{}
\item 
\end{list}
\vspace{-2em}
\noindent

%%%%%%%%%%%%%%%%%%%%%%%%%%%%%%%%%%%%%%%%%%%%%%%%%%
\section*{References}

{\scriptsize
\noindent
Altenhoff, W. J.\ \& Stumpff, P.\ 1995. Size estimate of "asteroid" 2060 chiron from 250GHz measurements. A\&A 293, L41-L42.

\noindent
Altenhoff, W. J., Menten, K. M.\ \& Bertoldi, F.\ 2001. Size determination of the Centaur Chariklo from millimeter-wavelength bolometer observations. A\&A 366, L9-L12.

\noindent
Altenhoff, W. J., Bertoldi, F.\ \& Menten, K. M.\ 2004. Size estimates of some optically bright KBOs. A\&A 415, 771-775.

\noindent
Aumann, H. H.\ \&  Walker, R. G.\ 1987. IRAS observations of the Pluto-Charon system. AJ 94, 1088-1091.

\noindent
Barr, A.C., Schwamb, M. E.\ 2016. Interpreting the densities of the Kuiper belt's dwarf planets. MNRAS 460, 1542-1548.

\noindent
Barucci, M. A., Merlin, F., Perna, D., Alvarez-Candal, A., M\"uller, T., Mommert, M.\ et al.\ 2012. The extra red plutino (55638) 2002 VE95. A\&A 539, A152, 6 pp.

\noindent
Batygin, K., Brown, M. E., \& Fraser, W. C.\ 2011. Retention of a Primordial Cold Classical Kuiper Belt in an Instability-Driven Model of Solar System Formation. ApJ 738, 13, 8pp.

\noindent
Bauer, J. M., Grav, T., Blauvelt, E., Mainzer, A. K., Masiero, J. R., Stevenson, R.\ et al.\ 2013. Centaurs and Scattered Disk Objects in the Thermal Infrared: Analysis of WISE/NEOWISE Observations. ApJ 773, 22, 11pp.

\noindent
Bertoldi, F., Altenhoff, W., Weiss, A., Menten, K. M.\ \&  Thum, C.\ 2006. The trans-neptunian object UB313 is larger than Pluto. Nature 439, 563-564.

\noindent
Bertrand, T. \& Forget, F. 2016. Observed glacier and volatile distribution on Pluto from atmosphere-topography processes. Nature 540, 86-89.

\noindent
Bowell, E., Hapke, B., Domingue, D., Lumme, K., Peltoniemi, J., Harris, A. W.\ 1989. Application of photometric models to asteroids. In: Asteroids II; Proceedings of the Conference, Tucson, AZ, Mar. 8-11, 1988. Univ. of Arizona Press, p524-556.

\noindent
Braga-Ribas, F., Sicardy, B., Ortiz, J. L., Snodgrass, C. Roques, F., Vieira-Martins, R.\ et al.\ 2014. A ring system detected around the Centaur (10199) Chariklo. Nature 508, 72-75.

\noindent
Brown, M. E., Trujillo, C.\ \&  Rabinowitz, D.\ 2004. Discovery of a Candidate Inner Oort Cloud Planetoid. ApJ 617, 645-649.

\noindent
Brown, M. E.\ \& Schaller, E. L.\ 2007. The Mass of Dwarf Planet Eris. Science 316, 1585.

\noindent
Brown, M. E., Schaller, E. L., Fraser, W. C.\ 2011. A Hypothesis for the Color Diversity of the Kuiper Belt. ApJ 739, L60, 5pp.

\noindent
Brown, M. E., Schaller, E. L., Fraser, W. C.\ 2012. Water Ice in the Kuiper Belt. AJ 143, 146, 7pp. 

\noindent
Brown, M. E.\ 2013. The Density of Mid-sized Kuiper Belt Object 2002 UX25 and the Formation of the Dwarf Planets. ApJL 778, L34, 5pp. 

\noindent
Brown, M. E.\ \& Butler, B. J.\ 2017. The Density of Mid-sized Kuiper Belt Objects from ALMA Thermal Observations. AJ 154, 19, 7pp.

\noindent
Brown, M. E.\ \& Butler, B. J.\ 2018. Medium-sized satellites of large Kuiper belt objects. arXiv:1801.07221.

\noindent
Brown, R. H.\ 1985. Ellipsoidal geometry in asteroid thermal models - The standard radiometric model. Icarus 64, 53-63.

\noindent
Brucker, M. J., Grundy, W. M., Stansberry, J. A., Spencer, J. R., Sheppard, S. S., Chiang, E. I., Buie, M. W.\ 2009. High albedos of low inclination Classical Kuiper belt objects. Icarus 201, 284-294.

\noindent
Butler, B. J., Gurwell, M., Lellouch, E., Moullet, A., Moreno, R., Bockelee-Morvan, D.\ et al.\ 2015. Long Wavelength Observations of Thermal Emission from Pluto and Charon with ALMA. DPS meeting \# 47, id.210.04.

\noindent
Buratti, B. J., Hicks, M. D., Dalba, P. A., Chu, D., O'Neill, A., Hillier, J. K.\ et al. 2015. Photometry of Pluto 2008-2014: Evidence of ongoing seasonal volatile transport and activity. ApJL 804, L6, 6pp.

\noindent
Buratti, B. J., Hofgartner, J. D., Hicks, M. D., Weaver, H. A., Stern, S. A., Momary, T.\ et al.\ 2017. Global albedos of Pluto and Charon from LORRI New Horizons observations. Icarus 287, 207-217.

\noindent
Campins, H., Telesco, C. M., Osip, D. J., Rieke, G. H., Rieke, M. J., Schulz, B.\ 1994. The color temperature of (2060) Chiron: A warm and small nucleus. AJ 108, 2318-2322.

\noindent
Cruikshank, D. P., Stansberry, J. A., Emery, J. P., Fern\'andez, Y. R., Werner, M. W., Trilling, D. E.\ et al.\ 2005. The High-Albedo Kuiper Belt Object (55565) 2002 AW197. ApJ 624, L53-L56.

\noindent
Cruikshank, D. P., Stansberry, J. A., Emery, J. P., van Cleve, J., Fern\'andez, Y. R., Werner, M. W.\ et al. 2006. Solar System Observations with Spitzer Space Telescope. in The Spitzer Space Telescope: New Views of the Cosmos, Proceedings ASP Conference Series 357, 9-12 November, 2004 in Pasadena, California, USA. Edited by L. Armus and W.T. Reach. San Francisco: Astronomical Society of the Pacific, 2006., p.23.

\noindent
Davies, J. R., Spencer, J., Sykes, M., Tholen, D., Green, S.\ 1993. (5145) Pholus. IAU Circ., No. 5698, \#2 (1993).

\noindent
Davies, J. R.\ 2000. Physical Characteristics of Trans-Neptunian Objects and Centaurs. in Minor Bodies in the Outer Solar System: Proceedings of the ESO Workshop Held at Garching, Germany, 2-5 Nov 1998. Edited by A. Fitzsimmons, D. Jewitt, and R.M. West. Springer-Verlag, 2000, p9.

\noindent
Delbo' M., Mueller, M., Emery, J. P., Rozitis, B., Capria, M. T. 2015. Asteroid Thermophysical Modeling. in Asteroids IV, Patrick Michel, Francesca E. DeMeo, and William F. Bottke (eds.), University of Arizona Press, Tucson, 895 pp. ISBN: 978-0-816-53213-1, p.107-128.

\noindent
Dias-Oliveira, A., Sicardy, B., Ortiz, J. L., Braga-Ribas, F., Leiva, R., Vieira-Martins, R.\ et al.\ 2017. Study of the Plutino Object (208996) 2003 AZ84 from Stellar Occultations: Size, Shape, and Topographic Features. AJ 154, 22, 13pp.

\noindent
Doressoundiram, A., Boehnhardt, H., Tegler, S. C., Trujillo, C.\ 2008. Color Properties and Trends of the Transneptunian Objects. in: The Solar System Beyond Neptune, ed. M. A. Barucci, H. Boehnhardt, D. Cruikshank, A. Morbidelli, University of Arizona Press, Tucson, 592pp, 91-104.

\noindent
Duffard, R., Pinilla-Alonso, N., Santos-Sanz, P., Vilenius, E., Ortiz, J. L., M\"uller, T.\ et al.\ 2014. TNOs are Cool: A Survey of the Transneptunian Region: A Herschel-PACS view of 16 Centaurs. A\&A 564, A92, 17 pp.

\noindent
Fern{\'a}ndez, Y. R., Jewitt, D. C.\ \&  Sheppard, S. S.\ 2002. Thermal Properties of Centaurs Asbolus and Chiron. AJ 123, 1050-1055.

\noindent
Ferrari, C.\ \& Lucas, A.\ 2016. Low thermal inertias of icy planetary surfaces. Evidence for amorphous ice? A\&A 588, A133, 14 pp.

\noindent
Fornasier, S., Lellouch, E., M\"uller, T., Santos-Sanz, P., Panuzzo, P., Kiss, C.\ et al.\ 2013. TNOs are Cool program: combined observations of 9 Centaurs and TNOs with PACS and SPIRE instruments onboard the Herschel space telescope. A\&A 555, A15, 22pp.

\noindent
Fornasier, S., Lazzaro, D., Alvarez-Candal, A., Snodgrass, C., Tozzi, G. P., Carvano, J. M.\ et al.\ 2014. The Centaur 10199 Chariklo: investigation into rotational period, absolute magnitude, and cometary activity. A\&A 568, L11, 5 pp.

\noindent
Fraser, W.C.\ \& Brown, M.E.\ 2012. The Hubble Wide Field Camera 3 Test of Surfaces in the Outer Solar System: The Compositional Classes of the Kuiper Belt. AJ 749, 33, 21 pp.

\noindent
Fraser, W.C., Brown, M.E., Morbidelli, A., Parker, A., Batygin, K.\ 2014. The Absolute Magnitude Distribution of Kuiper Belt Objects. ApJ 782, 100, 14 pp.

\noindent
Gerdes, D. W., Sako, M., Hamilton, S., Zhang, K., Khain, T., Becker, J. C.\ et al.\ 2017. Discovery and Physical Characterization of a Large Scattered Disk Object at 92 au. ApJ Letters 839, L15, 7 pp.

\noindent
Gladman, B., Holman, M., Grav, T., Kavelaars, J. Nicholson, P. Aksnes, K.\ et al.\ 2002. Evidence for an Extended Scattered Disk. Icarus 157, 269-279.

\noindent
Gladman, B., Marsden, B. G., VanLaerhoven, C.\ 2008. Nomenclature in the outer Solar System, in: The Solar System Beyond Neptune, ed. M. A. Barucci, H. Boehnhardt, D. Cruikshank, A. Morbidelli, University of Arizona Press, Tucson, 592pp, 43-57.

\noindent
Groussin, O., Lamy, P.\ \&  Jorda, L.\ 2004. Properties of the nuclei of Centaurs Chiron and Chariklo. A\&A 413, 1163-1175.

\noindent
Grundy, W. M., Noll, K. S., Stephens, D. C.\ 2005. Diverse albedos of small trans-neptunian objects. Icarus 176, 184-191.

\noindent
Grundy, W. M., Stansberry, J. A., Noll, K. S., Stephens, D. C., Trilling, D. E., Kern, S. D.\ et al. 2007. The orbit, mass, size, albedo, and density of (65489) Ceto/Phorcys: A tidally-evolved binary Centaur. Icarus 191, 286-297.

\noindent
Grundy, W. M., Noll, K. S., Virtanen, J., Muinonen, K., Kern, S. D., Stephens, D. C.\ et al.\ 2008. (42355) Typhon Echidna: Scheduling observations for binary orbit determination. Icarus 197, 260-268.

\noindent
Grundy, W. M., Cruikshank, D. P., Gladstone, G. R., Howett, C. J. A., Lauer T. R., Spencer, J. R.\ et al.\ 2016. The formation of Charon's red poles from seasonally cold-trapped volatiles. Nature 539, 65-68.

\noindent
Gulkis, S., Keihm, S., Kamp. L., Lee, S., Hartogh, P., Crovisier, J.\ et al.\ 2012. Continuum and spectroscopic observations of asteroid (21) Lutetia at millimeter and submillimeter wavelengths with the MIRO instrument on the Rosetta spacecraft. P\&SS 66, 31-42.

\noindent
Gundlach, B.\ \& Blum J.\ 2013. A new method to determine the grain size of planetary regolith. Icarus, 223 479-492.

\noindent
Hanu\v{s}, J., Delbo\', M., \v{D}urech, J., Al\'{i}-Lagoa, V.\ 2018. Thermophysical modeling of main-belt asteroids from WISE thermal data. Icarus, 309, 297-337.

\noindent
Harris, A. W.\ 1998. A Thermal Model for Near-Earth Asteroids. Icarus 131, 291-301.

\noindent
Hewison, T. J.\ \&  English, S. J.\ 1999. Airborne retrievals of snow and ice surface emissivity at millimeter wavelengths. IEEE Trans. Geosc. Remot. Sound.\ 37, 1871-1879.

\noindent
Howell, E., Marcialis, R., Cutri, R., Nolan, M., Lebofsky, L., Sykes, M.\ 1992. 1992 AD. IAU Circ., No. 5449, \#2 (1992). Edited by Green, D. W. E.

\noindent
Howett, C. J. A., Spencer, J. R., Pearl, J., Segura, M.\ 2010. Thermal inertia and bolometric Bond albedo values for Mimas, Enceladus, Tethys, Dione, Rhea and Iapetus as derived from Cassini/CIRS measurements. Icarus 206, 573-593.

\noindent
Janssen, M. A., Lorenz, R. D., West, R., Paganelli, F., Lopes, R. M., Kirk, R. L.\ et al.\ 2009. Titan's surface at 2.2-cm wavelength imaged by the Cassini RADAR radiometer: Calibration and first results. Icarus 200, 222-239.

\noindent
Jewitt, D.\ \& Luu, J.\ 1992. Submillimeter continuum observations of 2060 Chiron. AJ 104, 398-404.

\noindent
Jewitt, D.\ 1994. Heat from Pluto. AJ 107, 372-378.

\noindent
Jewitt, D.\ \& Kalas, P.\ 1998. Thermal Observations of Centaur 1997 CU26. ApJ 499, L103-L106.

\noindent
Jewitt, D., Aussel, H., Evans, A.\ 2001. The size and albedo of the Kuiper-belt object (20000) Varuna. Nature 411, 446-447.

\noindent
Jewitt, D.\ 2009. The Active Centaurs. AJ 137, 4296-4312.

\noindent
Keihm, S. J.\ 1984. Interpretation of the lunar microwave brightness temperature spectrum - Feasibility of orbital heat flow mapping. Icarus 60, 568-589.

\noindent
Keihm, S., Kamp, L., Gulkis, S., Hofstadter, M., Lee, S., Janssen, M., Choukroun, M.\ 2013. Reconciling main belt asteroid spectral flux density measurements with a self-consistent thermophysical model. Icarus 226, 1086-1102.

\noindent
Kiss, Cs., Szab\'o, G., Horner, J., Conn, B. C., M\"uller, T. G., Vilenius, E., Sarneczky, K.\ et al.\ 2013. A portrait of the extreme solar system object 2012 DR30. A\&A 555, A3, 13 pp.

\noindent
Kiss, C., P\'al, A., Farkas-Tak\'acs, A. I., Szab\'o, G. M., Szab\'o, R., Kiss, L. L., Moln\'ar, L.\ et al.\ 2016. Nereid from space: rotation, size and shape analysis from K2, Herschel and Spitzer observations. MNRAS 457, 2908-2917.

\noindent
Kiss, C., Marton, G., Farkas-Tak\'acs, A., Stansberry, J., M\"uller, T., Vink\'o, J.\ et al.\ 2017. Discovery of a satellite of the large trans-Neptunian object (225088) 2007 OR10. ApJL 838, L1, 5 pp.

\noindent
Kiss, Cs., Marton, G., Parker, A. H., Grundy, W., Farkas-Tak\'acs, A., Stansberry, J.\ et al.\ 2018. The mass and density of the dwarf planet (225088) 2007 OR 10. Icarus, submitted.

\noindent
Kovalenko, I. D., Doressoundiram, A., Lellouch, E., Vilenius, E., M\"uller, T., Stansberry, J., 2017. "TNOs are Cool": A survey of the trans-Neptunian region. XIII. Statistical analysis of multiple trans-Neptunian objects observed with Herschel Space Observatory. A\&A 608, A19, 8 pp.

\noindent
Lacerda, P., Jewitt, D., Peixinho, N.\ 2008. High-Precision Photometry of Extreme KBO 2003 EL61. AJ 135, 1749-1756.

\noindent
Lacerda, P., Fornasier, S., Lellouch, E., Kiss, C., Vilenius, E., Santos-Sanz, P.\ et al.\ 2014. TNOs Are Cool: A survey of the transneptunian region. XII. The albedo-color diversity of transneptunian objects. ApJL 793, L2, 6 pp.

\noindent
Lagerros, J. S. V. 1996. Thermal physics of asteroids. I. Effects of shape, heat conduction and beaming. A\&A 310, 1011-1020.

\noindent
Lagerros, J. S. V. 1997. Thermal physics of asteroids. III. Irregular shapes and albedo variegations. A\&A 325, 1226-1236.

\noindent
Lagerros, J. S. V. 1998. Thermal physics of asteroids. IV. Thermal infrared beaming. A\&A 332, 1123-1132.

\noindent
Lebofsky, L. A., Tholen, D. J., Rieke, G. H., Lebofsky, M. J.\ 1984. 2060 Chiron: Visual and thermal infrared observations. Icarus 60, 532-537.

\noindent
Le Gall, A., Leyrat, C., Janssen, M. A., Keihm, S., Wye, L. C., West, R.\ et al.\ 2014. Iapetus' near surface thermal emission modeled and constrained using Cassini RADAR Radiometer microwave observations. Icarus 241, 221-238.

\noindent
Leiva, R., Sicardy, B., Camargo, J.I.B., Ortiz, J. L., Desmars, J., B\'erard, D.\ et al.\ 2017. Size and Shape of Chariklo from Multi-epoch Stellar Occultations. AJ 154, 159, 23pp.

\noindent
Lellouch, E., Laureijs, R., Schmitt, B., Quirico, E., de Bergh, C., Crovisier, J., Coustenis, A.\ 2000a. Pluto's Non-isothermal Surface. Icarus 147, 220-250.

\noindent
Lellouch, E., Paubert, G., Moreno, R., Schmitt, B.\ 2000b. NOTE: Search for Variations in Pluto's Millimeter-Wave Emission. Icarus 147, 580-584.

\noindent
Lellouch, E., Moreno, R., Ortiz, J. L., Paubert, G., Doressoundiram, A., Peixinho, N.\ 2002. Coordinated thermal and optical observations of Trans-Neptunian object (20000) Varuna from Sierra Nevada. A\&A 391, 1133-1139.

\noindent
Lellouch, E., Kiss C., Santos-Sanz P., M\"uller T. G., Fornasier S., Groussin, O.\ et al.\ 2010. TNOs are cool: A survey of the trans-Neptunian region. II. The thermal lightcurve of (136108) Haumea. A\&A 518, L147, 5 pp.

\noindent
Lellouch, E., Stansberry, J., Emery, E., Grund, W., Cruikshank, D. P.\ 2011. Thermal properties of Pluto's and Charon's surfaces from Spitzer observations. Icarus 214, 701-716.

\noindent
Lellouch, E., Santos-Sanz, P., Lacerda, P., Mommert, M., Duffard, R., Ortiz, et al.\ 2013. "TNOs are Cool": A survey of the trans-Neptunian region. IX. Thermal properties of Kuiper belt objects and Centaurs from combined Herschel and Spitzer observations. A\&A 557, A60, 19pp.

\noindent
Lellouch, E., Santos-Sanz, P., Fornasier, S., Lim, T., Stansberry, J., Vilenius, E.\ et al.\ 2016. The long-wavelength thermal emission of the Pluto-Charon system from Herschel observations. Evidence for emissivity effect. A\&A 588, id.A2, 15 pp.

\noindent
Lellouch, E., Moreno, R., M\"uller, T., Fornasier, S., Santos-Sanz, P., Moullet, A.\ et al.\ 2017. The thermal emission of Centaurs and trans-Neptunian objects at millimeter wavelengths from ALMA observations. A\&A, 608, id. A45, 21 pp.
 
\noindent
Leyrat, C., Lorenz, R. D., Le Gall, A.\ 2016. Probing Pluto's underworld: Ice temperatures from microwave radiometry decoupled from surface conditions. Icarus 268, 50-55.

\noindent
Lim T. L., Stansberry J., M\"uller T. G., Mueller, M., Lellouch, E., Kiss, C.\ et al.\ 2010. TNOs are Cool: A survey of the trans-Neptuian region. III. Thermophysical properties of 90482 Orcus and 136472 Makemake. A\&A 518, L148, 5 pp.

\noindent
Lockwood, A. C., Brown, M. E., Stansberry, J.\ 2014. The Size and Shape of the Oblong Dwarf Planet Haumea. Earth, Moon and Planets 111, 127-137.

\noindent
Malhotra, R.\ 1995. The Origin of Pluto's Orbit: Implications for the Solar System Beyond Neptune. AJ 110, 420-429.

\noindent
Margot, J. L., Trujillo, C., Brown, M. E., Bertoldi, F.\ 2002. The size and albedo of KBO 2002 AW197. DPS Meeting \#34, id.17.03; BAAS 34, p.871.

\noindent
Marton, G., Kiss, Cs., Balog, Z., Lellouch, E., Vereb\'elyi, E., Klaas, U.\ 2015. Search for signatures of dust in the Pluto-Charon system using Herschel/PACS observations. A\&A 579, L9, 5 pp.

\noindent
Melita, M.D.\ \&  Licandro, J.\ 2012. Links between the dynamical evolution and the surface color of the Centaurs. A\&A 539, A144, 6 pp.

\noindent
Mishima, O., Klug, D. D., \& Whalley, E.\ 1983. The far-infrared spectrum of ice Ih in the range 8-25 cm$^{-1}$. Sound waves and difference bands, with application to Saturn's rings. J. Chem. Phys.\ 78, 6399-6404.

\noindent
Mitchell, D. L.\ \& de Pater, I.\ 1994. Microwave imaging of Mercury's thermal emission at wavelengths from 0.3 to 20.5 cm. Icarus 110, 2-32.

\noindent
Mommert, M., Harris, A. W., Kiss, C., P\'al, A., Santos-Sanz, P., Stansberry, J.\ et al.\ 2012. TNOs are cool: A survey of the trans-Neptunian region. V. Physical characterization of 18 Plutinos using Herschel-PACS observations. A\&A 541, A93, 17 pp.

\noindent
Mommert M.\ 2013. Remnant Planetesimals and their Collisional Fragments: Physical Characterization from Thermal Infrared Observations. Freie Universit\"at Berlin, PhD thesis, 300pp.

\noindent
Moullet, A., Lellouch, E., Moreno, R., Gurwell, M.\ 2011. Physical studies of Centaurs and Trans-Neptunian Objects with the Atacama Large Millimeter Array. Icarus 213, 382-392.

\noindent
M\"uller, T. G. \& Lagerros, J. S. V.\ 1998. Asteroids as far-infrared photometric standards for ISOPHOT. A\&A 338, 340-352.

\noindent
M\"uller, T. G. \& Lagerros, J. S. V.\ 2002. Asteroids as calibration standards in the thermal infrared for space observatories. A\&A 381, 324-339.

\noindent
M\"uller, T. G., Lellouch, E., B\"ohnhardt, H., Stansberry, J., Barucci, A., Crovisier, J.\ et al.\ 2009. TNOs are Cool: A Survey of the Transneptunian Region. EMP 105, 209-219.

\noindent
M\"uller T. G., Lellouch E., Stansberry J., Kiss, C., Santos-Sanz, P., Vilenius, E., et al.\ 2010. TNOs are Cool: A survey of the trans-Neptunian region. I. Results from the Herschel science demonstration phase (SDP). A\&A 518, L146, 5 pp.

\noindent
M\"uller, T., Kiss, Cs., Al{\'i}-Lagoa, V., Ortiz, J. L., Lellouch, E., Santos-Sanz, P.\ et al.\ 2018. Haumea's thermal emission revisited in the light of the occultation results. Icarus, submitted.

\noindent
Muhleman, D. O., \& Berge, G. L.\ 1991. Observations of Mars, Uranus, Neptune, Io, Europa, Ganymede, and Callisto at a wavelength of 2.66 MM. Icarus 92, 263-272.

\noindent
Myhrvold, N.\ 2018. Asteroid thermal modeling in the presence of reflected sunlight. Icarus 303, 91-113.

\noindent
Nimmo, F., Umurhan, O.; Lisse, C. M., Bierson, C. J., Lauer, T. R.; Buie, M. W.\ et al.\ 2017. Mean radius and shape of Pluto and Charon from New Horizons images. Icarus 287, 12-29.

\noindent
Noll, K. S., Grundy, W. M., Stephens, D. C., Levison, H. F., Kern, S. D.\ 2008. Evidence for two populations of classical transneptunian objects: The strong inclination dependence of classical binaries. Icarus 194, 758-768.

\noindent
Norwood, J., Hammel, H., Milam, S., Stansberry, J., Lunine, J. Chanover, N.\ et al.\ 2016. Solar System Observations with the James Webb Space Telescope. PASP 128:025004 (34pp).

\noindent
Ortiz, J. L., Sota, A., Moreno, R., Lellouch, E., Biver, N., Doressoundiram, A.\ et al.\ 2004. A study of Trans-Neptunian object 55636 (2002 TX300). A\&A 420, 383-388.

\noindent
Ortiz, J., Sicardy, B., Braga-Ribas, F., Alvarez-Candal, A., Lellouch, E., Duffard, R.\ et al.\ 2012. Albedo and atmospheric constraints of dwarf planet Makemake from a stellar occultation. Nature 491, 566-569.

\noindent
Ortiz, J. L., Duffard, R., Pinilla-Alonso, N., Alvarez-Candal, A., Santos-Sanz, P., Morales, N.\ et al.\ 2015. Possible ring material around centaur (2060) Chiron. A\&A 576, A18, 12pp.

\noindent
Ortiz, J. L., Santos-Sanz, P., Sicardy, B., Benedetti-Rossi, G., B\'erard, D., Morales, N.\ et al.\ 2017. The size, shape, density and ring of the dwarf planet Haumea from a stellar occultation. Nature 550, 219-223.

\noindent
Ostro, S. J., West, R. D., Janssen, M. A., Lorenz, R. D., Zebker, H. A., Black, G. J.\ et al.\ 2006. Cassini RADAR observations of Enceladus, Tethys, Dione, Rhea, Iapetus, Hyperion, and Phoebe. Icarus 183, 479-490.

\noindent
P\"atzold, M., Andert, T., Hahn, M., Asmar, S. W., Barriot, J.-P., Bird, M. K.\ 2016. A homogeneous nucleus for comet 67P/Churyumov-Gerasimenko from its gravity field. Nature 530, 63-65.  

\noindent
P\'al, A., Kiss, C., M\"uller, T. G., Santos-Sanz, P., Vilenius, E., Szalai, N.\ et al.\ 2012. "TNOs are Cool": A survey of the trans-Neptunian region. VII. Size and surface characteristics of (90377) Sedna and 2010 EK139. A\&A 541, L6, 4 pp.

\noindent
P\'al, A., Kiss, C., Horner, J., Szakats, R., Vilenius, E., M\"uller, T.\ et al.\ 2015. Physical properties of the extreme Centaur and super-comet candidate 2013 AZ60. A\&A 583, A93, 8 pp.

\noindent
P\'al, A., Kiss, C., M\"uller, T., Molnar, L., Szabo, R., Szabo, G. M.\ et al.\ 2016. Large Size and Slow Rotation of the Trans-Neptunian Object (225088) 2007 OR10. Discovered from Herschel and K2 Observations. AJ 151, 117, 8 pp.

\noindent
Pan, M.\ \& Sari, R.\ 2005. Shaping the Kuiper belt size distribution by shattering large but strengthless bodies. Icarus 173, 342-348.

\noindent
Parker, A., Pinilla-Alonso, N., Santos-Sanz, P., Stansberry, J., Alvarez-Candal, A., Bannister, M.\ et al.\ 2016a. Physical Characterization of TNOs with the James Webb Space Telescope. PASP 128:018010 (6pp).

\noindent
Parker, A. H., Buie, M. W., Grundy, W. M., Noll, K. S.\ 2016b. Discovery of a Makemakean Moon. ApJL 825, L9, 5 pp.

\noindent
Petit, J.-M., Kavelaars, J. J., Gladman, B. J., Jones, R. L., Parker, J. W., Van Laerhoven, C.\ et al.\ 2011. The Canada-France Ecliptic Plane Survey¿Full Data Release: The Orbital Structure of the Kuiper Belt. AJ 142, 131, 24pp.

\noindent
Preusker, F., Scholten, F., Matz, K.-D., Roatsch, T., Hviid, S. F., Mottola, S.\ et al.\ 2017. The global meter-level shape model of comet 67P/Churyumov-Gerasimenko. A\&A 607, L1, 5pp.

\noindent
Ragozzine, D.\ \& Brown, M. E.\ 2007. Candidate Members and Age Estimate of the Family of Kuiper Belt Object 2003 EL61. AJ 134, 2160-2167.

\noindent
Ries, P. A, \& Janssen, M.\ 2015. A large-scale anomaly in Enceladus' microwave emission. Icarus 257, 88-102.

\noindent
Romanishin, W.\ \&  Tegler, S.C.\ 2018. Albedos of Centaurs, Jovian Trojans, and Hildas. AJ 156, 19, 11 pp. 

\noindent
Santos-Sanz P., Lellouch E., Fornasier S., Kiss C., Pal A., M\"uller T. G.\ 2012. TNOs are Cool: A Survey of the Transneptunian Region IV. Size/albedo characterization of 15 scattered disk and detached objects observed with Herschel Space Observatory-PACS. A\&A 541, A92, 18 pp.

\noindent
Santos-Sanz, P., Lellouch, E., Groussin, O., Lacerda, P., M\"uller, T. G., Ortiz, J. L.\ et al.\ 2017. "TNOs are Cool": A survey of the trans-Neptunian region. XII. Thermal light curves of Haumea, 2003 VS2 and 2003 AZ84 with Herschel/PACS. A\&A 604, A95, 19 pp.

\noindent
Schindler, K., Wolf, J., Bardecker, J., Olsen, A., M\"uller, T., Kiss, C.\ et al.\ 2017. Results from a triple chord stellar occultation and far-infrared photometry of the trans-Neptunian object (229762) 2007 UK126. A\&A, 600, A12, 16 pp.

\noindent
Sicardy, B., Ortiz, J. L., Assafin, M., Jehin, E., Maury, A., Lellouch, E.\ et al.\ 2011. A Pluto-like radius and a high albedo for the dwarf planet Eris from an occultation. Nature 478, 493-496.

\noindent
Sierks, H., Barbieri, C., Lamy, P. L., Rodrigo, R., Koschny, D., Rickman, H.\ et al.\ 2015. On the nucleus structure and activity of comet 67P/Churyumov-Gerasimenko. Science 347, aaa1044.

\noindent
Spencer, J. R., \& Moore, J. M.\ 1992. The influence of thermal inertia on temperatures and frost stability on Triton. Icarus 99, 261-272.

\noindent
Stansberry, J. A., Cruikshank, D. P., Grundy, W. M., Emery, J. P., Osip, D. J., Fernandez, Y. R.\ et al.\ 2004. Far-IR Photometry of Centaurs and Kuiper Belt Objects with Spitzer. DPS meeting \#36, id.43.01; BAAS, Vol. 36, p.1173.

\noindent
Stansberry, J. A., Grundy, W. M., Margot, J. L., Cruikshank, D. P., Emery, J. P., Rieke, G. H.\ et al.\ 2006. The Albedo, Size, and Density of Binary Kuiper Belt Object (47171) 1999 TC36. ApJ 643, 556-566.

\noindent
Stansberry, J., Grundy, W., Brown, M., Cruikshank, D., Spencer, J., Trilling, D., Margot, J.-L.\ 2008. Physical Properties of Kuiper Belt and Centaur Objects: Constraints form the Spitzer Space Telescope, in The solar system beyond Neptune, ed. M. A. Barucci, H. Boehnhardt, D. Cruikshank, A. Morbidelli, Univ. of Arizona Press, Tucson, 161-179.

\noindent
Stansberry, J. A., Grundy, W. M., Mueller, M., Benecchi, S. D., Rieke, G. H., Noll, K. S.\ et al.\ 2012. Physical properties of trans-neptunian binaries (120347) Salacia-Actaea and (42355) Typhon-Echidna. Icarus 219, 676-688.

\noindent
Stern, S. A., Weintraub, D. A., Festou, M. C.\ 1993. Evidence for a Low Surface Temperature on Pluto from Millimeter-Wave Thermal Emission Measurements. Science 261, 1713-1716.

\noindent
Stern, S. A., Grundy, W. M., McKinnon, W. B., Weaver, H. A., Young, L., A.\ 2018. The Pluto System After New Horizons. ARA\&A 56, 357-392.

\noindent
Sykes, M. V., Cutri, R. M., Lebofsky, L. A., Binzel, R.\ 1987. IRAS serendipitous survey observations of Pluto and Charon. Science 237, 1336-1340.

\noindent
Sykes, M. V.\ \& Walker, R. G.\ 1991. Constraints on the diameter and albedo of 2060 Chiron. Science 251, 777-780.

\noindent
Sykes, M. V.\ 1999. IRAS Survey-Mode Observations of Pluto-Charon. Icarus 142, 155-159.

\noindent
Tedesco, E. F., Veeder, G. J., Dunbar, R. S., Lebofsky, L. A.\ 1987. IRAS constraints on the sizes of Pluto and Charon. Nature 327, 127-129.

\noindent
Tegler, S.C., \& Romanishin, W.\ 2000. Extremely red Kuiper-belt objects in near-circular orbits beyond 40 AU. Nature 407, 979-981.

\noindent
Tegler, S. C., Romanishin, W., \& Consolmagno, G.\ 2016. Two Color Populations of Kuiper Belt and Centaur Objects and the Smaller Orbital Inclinations of Red Centaur Objects. AJ 152, 210, 13 pp.

\noindent
Th\'ebault, P.\ \&  Doressoudiram, A.\ 2003. Colors and collision rates within the Kuiper belt: problems with the collisional resurfacing scenario. Icarus 162, 27-37.

\noindent
Thomas, N., Eggers, S., Ip, W.-H., Lichtenberg, G., Fitzsimmons, A., Jorda, L.\ et al.\ 2000. Observations of the Trans-Neptunian Objects 1993 SC and 1996 TL66 with the Infrared Space Observatory. ApJ 534, 446-455.

\noindent
Verbiscer, A., Porter, S. B., Benecchi, S., Kavelaars, J. J., Weaver, H. A., Spencer, J. R.\ et al. 2018. Photometric properties of distant KBOs observed by New Horizons LORRI at moderate and high phase angles. The Transneptunian Solar System, March 26-29, 2018, Coimbra, Portugal.

\noindent
Vilenius, E., Kiss, C., Mommert, M., M\"uller, T., Santos-Sanz, P., P\'al, A.\ et al.\ 2012. TNOs are cool: A survey of the trans-Neptunian region. VI. Herschel/PACS observations and thermal modelling of 19 classical Kuiper Belt objects. A\&A 541, A94, 17 pp.

\noindent
Vilenius, E., Kiss, C., M\"uller, T., Mommert, M., Santos-Sanz, P., P\'al, A.\ et al.\ 2014. "TNOs are Cool": A survey of the trans-Neptunian region. X. Analysis of classical Kuiper belt objects from Herschel and Spitzer observations. A\&A 564, A35, 18 pp.

\noindent
Vilenius, E., Stansberry, J., M\"uller, T., Mueller, M., Kiss, C., Santos-Sanz, P.\ et al.\ 2018. "TNOs are Cool": A survey of the trans-Neptunian region. XIV. Size/albedo characterization of the Haumea family observed with Herschel and Spitzer. A\&A in press, 15 pp.

\noindent
Volk, K.\ \&  Malhotra, R.\ 2008. The Scattered Disk as the Source of the Jupiter Family Comets. ApJ 687, 714-725.

\noindent
Young, L.A.\ 2013. Pluto's Seasons: New Predictions for New Horizons. ApJL 766, L22, 6 pp.

\noindent
Wolters, S. D.\ \&  Green, S. F.\ 2009. Investigation of systematic bias in radiometric diameter determination of near-Earth asteroids: the night emission simulated thermal model (NESTM). MNRAS 400, 204-218.

\noindent
Wong, I.\ \&  Brown, M. E.\ 2016. A Hypothesis for the Color Bimodality of Jupiter Trojans. AJ 152, 90, 8 pp. 

}
%%%%%%%%%%%%%%%%%%%%%%%%%%%%%%%%%%%%%%%%%%%%%%%%%%

\end{document}